\documentclass[aps,prd,floatfix,showpacs,nofootinbib,twoside,notitlepage,11pt]{revtex4-1}
\usepackage{amssymb}
\usepackage{amsmath}
\usepackage{amstext}
\usepackage{graphicx,epsfig}
\usepackage{verbatim} 
\usepackage{fancybox}
\usepackage{color}
\usepackage{ulem}
\usepackage{enumitem}
\usepackage{subfigure}
\usepackage{bbm}
\usepackage{dsfont}
\usepackage{amsmath}	%Allows \begin{equation},\begin{align},\underbrace{2+2=4}_{\text{obvious}}
\usepackage{bm}		%Allows {\bm \nabla \alpha} for bold symbold
\usepackage{amssymb}	%Allows use math symbols
\usepackage{color} 	%Allows {\color{red} Hello World} or \colorbox{blue}{Hello World}
\usepackage{graphicx} 	%Allows \includegraphics{figure.pdf}
\usepackage[T1]{fontenc}
\usepackage{mathrsfs} 	%Allows \mathscr{HELLO}
\usepackage[sort&compress]{natbib} %Concatenate [1,2,3] -> [1-3]
\usepackage[colorlinks=true]{hyperref}
\usepackage{hyperref} 	%Automatically links \label and \ref commands; Always load last
\usepackage[all]{hypcap}%Ensures figure hyperlinks are positioned to display entire figure
\hypersetup{
 	citecolor=blue,		%color of links to bibliography
    	urlcolor=blue		%color of external links
	    }
\newcommand{\xvec}{{\bm x}}
\newcommand{\Avec}{{\bm A}}
\newcommand{\kvec}{{\bm k}}
\newcommand{\Mol}{\text{\rm M\o l}}
\newcommand{\X}{{\text{\sc x}}}
\newcommand{\DM}{{\text{\sc dm}}}
\newcommand{\RH}{{\text{\sc rh}}}
\newcommand{\BD}{\text{\sc bd}}
\newcommand{\Lcal}{\mathcal{L}}
\newcommand{\Mcal}{\mathcal{M}}
\newcommand{\Scal}{\mathcal{S}}
\newcommand{\Mpl}{M_{\rm pl}}
\newcommand{\GeV}{\ \mathrm{GeV}}

\newcommand{\Zbb}{\mathbb{Z}}
\newcommand{\sref}[1]{Sec.~\ref{#1}}
\newcommand{\eref}[1]{Eq.~(\ref{#1})}
\newcommand{\rref}[1]{Ref.~\cite{#1}}
\newcommand{\ud}{\mathrm{d}}
\newcommand{\per}{\ . \ }
\newcommand{\fref}[1]{Fig.~\ref{#1}}
\newcommand{\com}{\ , \ }
\newcommand{\pref}[1]{(\ref{#1})}

\newcommand{\aref}[1]{Appendix~\ref{#1}}
\newcommand{\erefs}[2]{Eqs.~(\ref{#1})~and~(\ref{#2})}

\newcommand{\nn}{\nonumber \\}

\begin{document}

\setlength{\pdfpageheight}{\paperheight}
\setlength{\pdfpagewidth}{\paperwidth}

%------------------------------
%  Title Page
%------------------------------ 

\title{Superheavy dark matter through Higgs portal operators}

\author{Edward W. Kolb and Andrew J. Long}

\affiliation{Kavli Institute for Cosmological Physics and the Enrico Fermi Institute, The University of Chicago, 5640 S. Ellis Ave, Chicago, IL 60637, USA}

\begin{abstract}
The WIMPzilla hypothesis is that the dark matter is a super-weakly-interacting and superheavy particle.  Conventionally, the WIMPzilla abundance is set by gravitational particle production during or at the end of inflation.  In this study we allow the WIMPzilla to interact directly with Standard Model fields through the Higgs portal, and we calculate the thermal production (freeze-in) of WIMPzilla dark matter from the annihilation of Higgs boson pairs in the plasma.  The two particle-physics model parameters are the WIMPzilla mass and the Higgs-WIMPzilla coupling.  The two cosmological parameters are the reheating temperature and the expansion rate of the universe at the end of inflation.  We delineate the regions of parameter space where either gravitational or thermal production is dominant, and within those regions we identify the parameters that predict the observed dark matter relic abundance.  Allowing for thermal production opens up the parameter space, even for Planck-suppressed Higgs-WIMPzilla interactions.  
\end{abstract}

%Cosmology, 98.80.-k
%Dark matter, 95.35.+d
%quantum fields in curved spacetime, 04.62.+v
%standard model Higgs boson, 14.80.Bn
%Higgs sector extensions, 12.60.Fr

\date{\today}

\maketitle

%==================================
% Introduction
\section{Introduction}\label{sec:Introduction}
%==================================

A superheavy particle, also known as the WIMPzilla, is an attractive dark matter (DM) candidate.  In general the WIMPzilla mass can be much larger than the weak scale, and possibly as large as the Hubble scale during inflation.  In this study we will consider WIMPzilla masses greater than about $10^8\GeV$. Unlike the lower-mass WIMP dark matter candidates, thermal freeze-out in a radiation dominated universe would dramatically overpredict the relic abundance of WIMPzilla dark matter \cite{Griest:1989wd}.  Consequently, the WIMPzilla is assumed to have very weak (or possibly nonexistent) interactions with particles in the plasma.  Various nonthermal production mechanisms have been explored for noninteracting WIMPzillas \cite{Chung:1998ua,Chung:1999ve,Fedderke:2014ura}.  Perhaps the most elegant of these mechanisms is the gravitational production of WIMPzillas during, or at the end of, inflation \cite{Chung:1998zb,Kuzmin:1998kk}.  This scenario only requires the WIMPzilla to couple to gravity; specifically, there need not be any direct coupling with the Standard Model (SM) fields or the inflaton. 

However, from an effective field theory (EFT) perspective we also expect the WIMPzilla to have interactions with SM fields suppressed by the scale of new physics, which may be as high as the Planck scale.  The leading-order interactions should be with the SM Higgs field through the so-called Higgs portal \cite{Silveira:1985rk,Burgess:2000yq,Patt:2006fw}.  In this article, we explore the consequences of these Higgs-WIMPzilla interactions.  

We suppose that heavy states at the scale of new physics mediate interactions between the WIMPzilla and the SM particles, and at the energy scales for thermal production these interactions can be described by an effective field theory.  The WIMPzilla-SM couplings take a different form depending on the spin of the WIMPzilla particle.  We will consider three cases: 1) a spin-0 WIMPzilla represented by the real scalar field $\phi(x)$; 2) a spin-1/2 WIMPzilla represented by the Majorana spinor field $\psi(x)$; and 3) a spin-1 WIMPzilla represented by a real vector field $A(x)$. 

Henceforth, when discussing WIMPzillas in this paper, ``scalar'' will refer to a real massive field with one scalar degree of freedom, ``fermion'' will refer to a massive Majorana fermion with two degrees of freedom, and ``vector'' will refer to a massive real vector field with three degrees of freedom.

The Higgs-squared operator $\Phi^\dagger\Phi$ is the only dimension-two SM operator that is Lorentz invariant and gauge invariant.  Thus, it is reasonable to expect that the leading-order WIMPzilla--SM interaction is through the Higgs field, denoted as $\Phi(x)$.  The interactions contribute terms to the Lagrangian of the form 
\begin{equation}\label{eq:interactions}
- \Lcal_{\rm int} \supseteq \left\{
	\begin{array}{ll}
	\dfrac{\kappa_\phi}{2} \ \phi^2 \Phi^\dagger \Phi & \quad \phi = \mathrm{scalar\ WIMPzilla} \\ [1em]
	\dfrac{\kappa_\psi}{2} \ \dfrac{1}{\Mpl} \ \psi\psi \Phi^\dagger \Phi & \quad \psi = \mathrm{fermion\ WIMPzilla} \\ [1em]
	\dfrac{\kappa_A}{2} \ \dfrac{m^2}{\Mpl^2} \ g^{\mu \nu} A_\mu A_\nu \Phi^\dagger \Phi & \quad A_\mu = \mathrm{vector\ WIMPzilla}
	\end{array}
\right.
\end{equation}
where $g^{\mu \nu}$ is the inverse metric and $m$ is the WIMPzilla mass (we will use $m$ to denote the WIMPzilla mass for all three models).  We use the reduced Planck mass $\Mpl \equiv \sqrt{1/8\pi G} \simeq 2.43 \times 10^{18} \GeV$ to normalize the irrelevant operators.  If the scale of new physics is lower than $\Mpl$, then this would correspond to $|\kappa| > 1$. To ensure that the WIMPzilla is a stable dark matter candidate, we enforce a $\Zbb_2$ symmetry on the WIMPzilla field.  This forbids operators such as the neutrino portal $L \Phi \psi$ and gauge-kinetic mixing terms $\partial_\mu A_{\nu} B^{\mu \nu}$, where $L$ is the SM lepton doublet and $B^{\mu\nu}$ is formed from SM gauge fields.  For the vector WIMPzilla coupling we include a factor of $m^2/\Mpl^2$ because this gauge-noninvariant operator must vanish in the limit $m \to 0$ where the gauge symmetry is restored.  

In this paper we study the range in parameter space where the number of WIMPzillas produced after inflation through the Higgs portal operators would exceed the number of WIMPzillas produced through gravitational processes.   The interactions in \eref{eq:interactions} allow WIMPzillas to be pair produced from annihilations of Higgs/anti-Higgs pairs in the plasma (prior to electroweak symmetry breaking).  In this way the DM abundance would be set by freeze-in, like gravitino DM \cite{Ellis:1984eq,Kawasaki:1994af} (for a recent review, see \rref{Bernal:2017kxu}).  We will also study where production through the Higgs portal dominates and produces WIMPzillas in the correct abundance to be dark matter.

The remainder of the article is organized as follows.  In \sref{sec:Interactions} we briefly motivate the Higgs-WIMPzilla interactions that appear in \eref{eq:interactions}.  In \sref{sec:Gravitational} we review how WIMPzilla dark matter can be generated from gravitational particle production.  The new work primarily appears in \sref{sec:Thermal} where we calculate the abundance of WIMPzilla dark matter that is produced thermally from interactions with the Higgs as described in \eref{eq:interactions}.  Our main results are summarized in \sref{sec:ThermalDM}, in which we illustrate the regions of parameter space where the current dark-matter relic abundance can be explained by either gravitational or thermal production of superheavy particles.  We conclude in \sref{sec:Conclusion} and suggest directions for future work.

%==================================
% Higgs-WIMPzilla Interactions
\section{Higgs-WIMPzilla Interactions}\label{sec:Interactions}
%==================================

In the spirit of effective field theory, there is nothing to prevent us from writing the operators that appear in \eref{eq:interactions}.  In fact, these operators are simply the product of the separate WIMPzilla and Higgs mass operators, and therefore any symmetry that forbids the interaction operators must also forbid the separate mass operators.  Then, for a massive WIMPzilla, one generally expects the Higgs-WIMPzilla interaction to be present.  Nevertheless, it is instructive to give examples of how such operators might originate from new physics at a higher mass scale.  (Here ``higher mass'' refers to masses larger than the expansion rate during inflation, the mass of the WIMPzilla, or the temperature of the universe during thermal production.)  

First consider the scalar WIMPzilla operator. The leading-order term coupling scalar WIMPzillas to the SM Higgs is dimension-four, so there is not an explicit suppression by the scale of new physics.  However there might be a suppression encoded in $\kappa_\phi$.  To see how this might arise, imagine that there is an approximate shift symmetry that forbids the term $\phi^2\Phi^\dagger\Phi$, but the shift symmetry is explicitly broken by dimension-six operators $\Lambda^{-2}\chi^4\phi^2$ and $\Lambda^{-2}\phi^2 \chi^2\Phi^\dagger\Phi$ involving a new scalar field $\chi$ and a scale of new physics $\Lambda$. If $\chi$ obtains a vacuum expectation $\langle\chi\rangle$, the first term would generate a mass $\langle\chi\rangle^2/\Lambda$ for the $\phi$, and the second term would result in a term of the form $(\langle\chi\rangle^2/\Lambda^2)\phi^2\Phi^\dagger\Phi$ coupling the WIMPzilla to the Higgs.  If we identify $\kappa_\phi/2=(\langle\chi\rangle/\Lambda)^2$,  then the scale of new physics $\Lambda$ would appear implicitly in $\kappa_\phi$, and $|\kappa_\phi|$ could be much less than unity.  On the other hand, one can imagine that there is just a $\kappa_\phi\phi^2\Phi^\dagger\Phi$ coupling where $|\kappa_\phi|$ is of order unity.   

In fact, from the EFT perspective, we should also allow the WIMPzilla to couple directly to the inflaton field \cite{Kannike:2016jfs}.  However, the argument above can be applied to explain why the inflaton-WIMPzilla coupling might be small.   More generally, a direct WIMPzilla-inflaton coupling opens a new channel for nonthermal WIMPzilla production in which the latter is produced directly from the decay of the inflaton during reheating or from a parametric resonance during preheating \cite{Kolb:1996jt,Chung:1998rq}.  We do not consider these additional WIMPzilla production mechanisms in this work.  

Now consider fermionic WIMPzillas.  Imagine a UV-complete model with the WIMPzilla $\psi$ and, again, a new scalar field $\chi$ of mass $\Lambda$, with interaction terms $g\psi \psi \chi$ and $\mu \chi \Phi^\dagger\Phi$, where $g$ is a Yukawa-type dimensionless coupling and $\mu$ has mass dimension one.  At scales much less than $\Lambda$, integrating out the $\chi$ field generates an effective term $(g\mu/\Lambda^2)\psi\psi\Phi^\dagger\Phi$.  We can then identify $\kappa_\psi/2\Mpl = g\mu/\Lambda^2$.

Finally, consider a possibility for vector WIMPzillas.  A mass for the vector field breaks gauge symmetry, so it is natural to imagine that it arises through a Higgs mechanism.  Consider the UV theory to include a scalar field $\chi$ charged under the gauged U$(1)$.  The covariant derivative of $\chi$ is $D_\mu\chi-igA_\mu\chi$, and the kinetic term for $\chi$, $D_\mu\chi D^\mu\chi^*$ generates a term $g^2A^\mu A_\mu\chi\chi^*$.  When $\chi$ develops a vacuum expectation value, a mass of $m^2=g^2\langle\chi\rangle^2$ for the vector field is generated.  Now if the Higgs field is coupled to $\chi$ through a term $\Lambda^{-2}D_\mu\chi D^\mu\chi^*\Phi^\dagger\Phi$, we would have a term $g^2\Lambda^{-2}A_\mu A^\mu \chi\chi^*\Phi^\dagger\Phi$. When $\chi$ gets a vacuum expectation value, the term becomes $(m/\Lambda)^2 A_\mu A^\mu \Phi^\dagger\Phi$.  So we would identify $\kappa_A/2\Mpl^2= \Lambda^{-2}$.

These examples are not meant to be the simplest nor most elegant UV completions, but they serve to illustrate how the terms in \eref{eq:interactions} might plausibly arise.  Moreover, this exercise lets us estimate what might serve as a reasonable range of values for the magnitude of the coefficients $\kappa_\phi$, $\kappa_\psi/\Mpl$, or $\kappa_A m^2/\Mpl^2$.  For the scalar model, $\kappa_\phi\phi^2\Phi^\dagger \Phi$ is a mass-dimension-four operator, and perturbative unitarity requires $|\kappa_\phi|<4\pi$.  For the fermion model, the Higgs portal interaction $(\kappa_\psi / M_{\rm pl}) \psi \psi \Phi^\dagger \Phi$ is nonrenormalizable, and the effective field theory is only reliable at energy scales that are small compared to the cutoff $\Mpl / |\kappa_\psi|$. In \sref{sec:Thermal} we will see that WIMPzillas are produced at a temperature (i.e., an energy scale) of $T_{\rm max}$, which is the maximum temperature of the universe after inflation.  Therefore, the validity of the EFT requires $|\kappa_\psi|/\Mpl<1/T_{\rm max}$ in the fermion model, and a similar argument in the vector WIMPzilla model leads to $|\kappa_A|/\Mpl^2<1/T^2_{\rm max}$.  So as a rough limit, we take 
\begin{align}\label{eq:kappamax}
	|\kappa_\phi| < 10^1 \, , \quad 
	|\kappa_\psi| < \dfrac{\Mpl}{T_{\rm max}} \sim 10^6 \, , \quad \text{and} \quad 
	|\kappa_A| < \dfrac{\Mpl^2}{T_{\rm max}^2} \sim 2 \times10^{12} \com
\end{align}
where we have used \eref{eq:Tmax} to evaluate $T_{\rm max} \sim 2 \times 10^{12} \GeV$ for the fiducial parameters.

%==================================
% Gravitational Production of WIMPzillas 
\section{Gravitational Production of WIMPzillas}\label{sec:Gravitational}
%==================================

If conformal invariance is not respected in the expanding universe, fields develop an effectively time-dependent dispersion relation due to their coupling with gravity \cite{Parker:1969au}.  If the dispersion relation evolves nonadiabatically for some Fourier modes, then the field is excited out of its vacuum state, which corresponds to particle production \cite{SCHRODINGER1939899}.  This phenomenon is similar to the behavior of a simple quantum harmonic oscillator when the spring constant changes abruptly.  An ideal environment for gravitational particle production is the transition from the accelerated expansion of cosmological inflation into the decelerated expansion of a matter- or radiation dominated universe \cite{Ford:1986sy,Yajnik:1990un}.  

Various people have studied the gravitational production of superheavy dark matter during inflation.  Since gravitational particle production can occur even when the particle in question has only a minimal gravitational interaction, the model is fully determined by specifying the particle's mass and spin.  The authors of Refs.~\cite{Chung:1998zb,Kuzmin:1998kk,Fedderke:2014ura} studied the gravitational production of scalar (spin-0) dark matter, those of Refs.~\cite{Kuzmin:1998kk,Chung:2011ck} studied spin-1/2 fermion dark matter, and those of Refs.~\cite{Dimopoulos:2006ms,Graham:2015rva} studied the vector (spin-1) dark matter case.  Whereas the scalar and fermion studies focused on superheavy (WIMPzilla) dark matter, the vector studies focused instead on superlight dark matter.  We are not aware of any studies of gravitational particle production with spin-3/2 fermions, and we do not consider that possibility further here.  

In order to determine the spectrum and relic abundance of dark matter that is produced by gravitational particle production, one can perform the following calculation.  First, one specifies a model of inflation and reheating, which fixes the evolution of the spacetime background.  The metric is in the Friedmann-Robertston-Walker (FRW) form with scale factor $a(t)$ and Hubble parameter $H(t) = \dot{a}/a$ at time $t$.  Next, one derives the dark matter field equation and solves it assuming the Bunch-Davies initial condition.  From the late-time behavior, one extracts the Bogoliubov coefficient, which is denoted as $\beta(\kvec)$ for the Fourier mode with comoving momentum $\kvec$.  Finally, one calculates the physical number density of dark matter particles at late times as 
\begin{align}\label{eq:n_Bogo}
n(t) = \frac{g}{a^3(t)} \int \! \! \frac{\ud^3 \kvec}{(2\pi)^3} \, |\beta({\kvec})|^2   \per
\end{align}
The factor $g$ counts the internal spin and flavor degrees of freedom: $g = 1$ for a scalar, $g = 2$ for a  fermion, $g = 2$ for the transverse polarizations of a vector, and $g = 1$ for the longitudinal polarization.  Equation~(\ref{eq:n_Bogo}) assumes that the dark matter does not participate in any particle-number-changing reactions after production, and consequently the comoving number density $a^3 n$ is conserved.  

\begin{figure}[t]
\begin{center}
\includegraphics[width=0.75\textwidth]{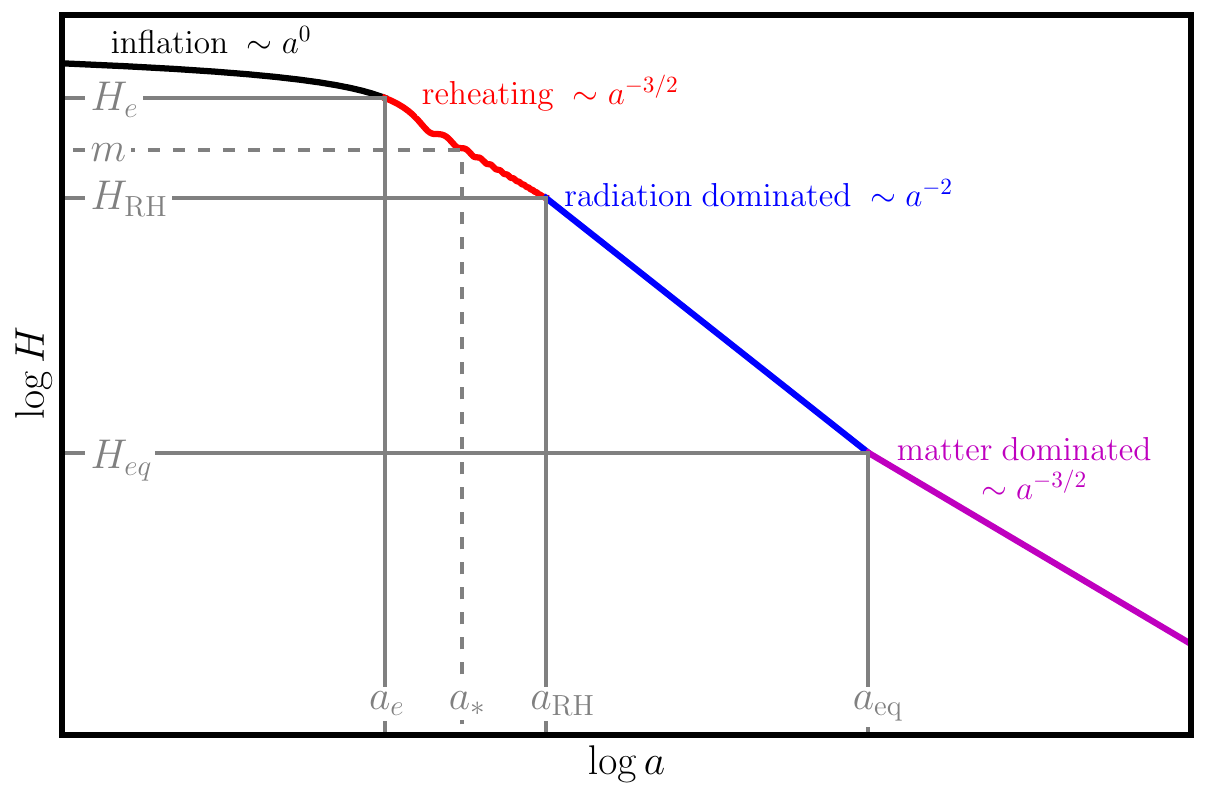}
\caption{\label{fig:H_of_a}
An illustration of the background cosmology assumed in this work.  On a log-log scale, we show how the Hubble parameter $H(t)$ varies with the monotonically growing FRW scale factor $a(t)$.  The WIMPzilla mass $m$ defines a time $t_\ast$ such that $H_\ast \equiv H(t_\ast) = m$ and $a_\ast \equiv a(t_\ast)$.  The very recent epoch of accelerated expansion is not shown.
}
\end{center}
\end{figure}

In each of the three dark matter models we assume the same background spacetime evolution, which is illustrated schematically in \fref{fig:H_of_a}.  Initially an epoch of inflation drives the accelerated expansion of the universe.  For concreteness we assume an inflaton potential quadratic in the inflaton field (chaotic inflation); we do not expect that our results will depend sensitively on this assumption \cite{Chung:2001cb}.  We assume an inflaton mass of $2 \times 10^{13} \GeV$, so  the Hubble parameter at the end of inflation is $H_e \equiv H(t_e) \simeq 10^{13} \GeV$, and 60 e-foldings before the end of inflation the expansion rate is $H_{\rm inf} \simeq 10^{14} \GeV$.  We define the end of inflation by $\ddot{a}(t_e) = 0$, which corresponds to the time when the comoving Hubble radius ($1/aH = 1/\dot{a}$) begins to grow.  Inflation is followed by an epoch of reheating during which time the inflaton field oscillates about the minimum of its potential and the universe is effectively matter dominated.  We assume that the plasma is generated by the perturbative decay of the inflaton.  If $\Gamma$ denotes the inflaton decay width, then reheating is approximately completed at time $t = t_\RH$ when $H_\RH \equiv H(t_\RH) \approx \Gamma$.  At this time the energy density of the plasma exceeds the energy in the coherent inflaton oscillations, and we say that the plasma has reached the reheat temperature, denoted by $T_\RH$, which we take as a free parameter.  However, the maximum temperature during reheating, $T_{\rm max}$, will generally exceed $T_\RH$ \cite{Giudice:2000ex}, and this fact is important for our study of thermal WIMPzilla production in \sref{sec:Thermal}.  Subsequently, we have a standard big bang cosmology: reheating is followed by an epoch of radiation domination that lasts until the dark matter energy density comes to dominate and heralds the epoch of WIMPzilla domination (dark matter domination), which approximately lasts until today when dark energy dominates.  

Finally let us remark that it is only meaningful to talk about the number density of gravitationally produced particles \eref{eq:n_Bogo} at late times.  There is a time $t = t_\ast$ at which the Hubble parameter $H$ decreases below $m$ such that $H_\ast \equiv H(t_\ast) = m$.  After this time, all of the (nonrelativistic) Fourier modes of the WIMPzilla field will be oscillating and their amplitudes will decay such that the energy density of the WIMPzilla field redshifts like pressureless dust, namely $\rho \sim a^{-3}$, and then we define $n = \rho/m$.  By this time, the evolution of $\omega_k$ has become adiabatic, and $|\beta(\kvec)|^2$ in \eref{eq:n_Bogo} is well defined.  We assume that the time $t_\ast$ occurs while the universe is still in the matter dominated phase of reheating.  Thus we focus on larger WIMPzilla masses that satisfy $m > H_\RH \simeq (1.4 \times 10^8 \GeV) (T_\RH / 10^{13} \GeV)^2 \sqrt{g_\ast(T_\RH)/106.75}$.  

In the following subsections we provide additional details of the gravitational particle production calculation for the scalar, fermion, and vector dark matter models, and we summarize the salient results that are relevant to our analysis.  

%=============
%=============

%==================================
% Gravitational WIMPzilla Production Scalars
\subsection{Scalars}\label{sec:Gravitational_Scalars}
%==================================

Consider a scalar field $\phi(x)$ with a nonminimal gravitational interaction.  The action for this field is given by
\begin{equation}\label{eq:RS_action_1}
S\left[\phi(x),g_{\mu \nu}(x)\right] = \int \! \ud^4 x \, \sqrt{-g} \, \left[ \frac{1}{2} g^{\mu \nu} \partial_{\mu} \phi \partial_{\nu} \phi - \frac{1}{2} m^2 \phi^2 - \frac{1}{2}\xi R\phi^2 - \frac{1}{2} \Mpl^2 R \right] \com
\end{equation}
where $R$ is the Ricci scalar and $\xi$ is a dimensionless coupling.  Two well-studied choices for $\xi$ are $\xi=0$ (``minimal'' coupling) and $\xi=1/6$ (``conformal'' coupling).  Note that $\xi=0$ is not particularly special, because even if $\xi=0$ at some energy scale, quantum corrections would induce $\xi \neq 0$ at other energy scales \cite{Nelson:1982kt,Shapiro:2015ova,Espinosa:2015qea}.  On the other hand, $\xi=1/6$ is a quasi-fixed point of the renormalization-group flow because the theory enjoys an approximate conformal symmetry in the limit $\xi \to 1/6$, which is spoiled by the mass parameter $m^2 / M_{\rm pl}^2 \neq 0$, as well as possibly the nongravitational interactions of $\phi$.  That said, we will analyze gravitational particle production for $\xi=0$ and $\xi=1/6$, and comment on the quantitative changes for other values of $\xi$.  Presumably, some theory around the Planck scale will determine a value of $\xi$.  Running $\xi$ down to an energy scale of $H_{\rm inf}$ would not change the value very much.

In an FRW spacetime, the field operator may be decomposed into mode functions $\chi_k(\eta)$, which only depend on the conformal time $\eta$ [$\ud\eta = \ud t/a(t)$] and the modulus of the comoving wave vector $|\kvec| \equiv k$ (owing to the homogeneity and isotropy of the background metric).  The mode decomposition is written as\footnote{The factor of $a^{-1}(\eta)$ ensures the field and its conjugate momentum satisfy the usual algebra $\left[\phi(\eta,\xvec) \, , \, \phi(\eta,\xvec^{\prime})\right] = \left[\pi(\eta,\xvec) \, , \, \pi(\eta,\xvec^{\prime})\right] = 0$ and $\left[\phi(\eta,\xvec) \, , \, \pi(\eta,\xvec^{\prime})\right] = - \left[\pi(\eta,\xvec) \, , \, \phi(\eta,\xvec^{\prime})\right] = i \, \delta^{(3)}(\xvec - \xvec^{\prime})$.}
\begin{align}\label{eq:RS_mode_decomp}
\phi(\eta,\xvec) = \frac{1}{a(\eta)} \, \int \! \! \frac{\ud^3 \kvec}{(2\pi)^3} \, \left[ \hat{a}(\kvec) \, \chi_k(\eta) \, e^{i \kvec \cdot \xvec} + \hat{a}^\dagger(\kvec)\, \chi_k^\ast (\eta)\, e^{-i \kvec \cdot \xvec} \right] \com
\end{align}
where $\hat{a}^\dagger(\kvec)$ and $\hat{a}(\kvec)$ are the creation and annihilation operators.   The mode functions satisfy the wave equation 
\begin{equation}\label{eq:RS_mode_eqn_chi}
\partial_{\eta}^2 \chi_k(\eta) + \omega_k^2(\eta) \, \chi_k(\eta)= 0  \com
\end{equation}
where the dispersion relation is 
\begin{align}\label{eq:dispersion}
\omega_k^2(\eta) = k^2 + a^2 m^2  - (1/6 - \xi) \, a^2 R \com 
\end{align}
and $R(\eta) = 12 H^2(\eta) + 6 \, a^{-1}(\eta)\, \partial_\eta H(\eta)$ in an FRW spacetime.  If the matter driving the FRW expansion has an effective equation of state $w = p / \rho$, the deceleration equation is written as $a\partial_\eta H = -(3/2) a^2 H^2 (1+w)$.  Then the dispersion relation \eref{eq:dispersion} during inflation ($w \approx -1$), matter domination ($w \approx 0$), and radiation domination ($w \approx 1/3$) is given by 
\begin{equation}\label{eq:dispersions}
\omega_k^2(\eta) = \left\{ \begin{array}{ll}
k^2 + a^2 m^2  - 2 (1 - 6 \xi) a^2 H^2            & \quad \mathrm{during\ inflation} \\ [1em]
k^2 + a^2 m^2  - \dfrac{1}{2} (1 - 6 \xi) a^2 H^2 & \quad \mathrm{during\ matter dominated\ expansion} \\ [1em]
k^2 + a^2 m^2                                     & \quad \mathrm{during\ radiation dominated\ expansion} \per
\end{array}\right.
\end{equation}
We will discuss below how particle production results from the nonadiabatic evolution of $\omega_k^2(\eta)$ during cosmological expansion.  The dispersion relation \eref{eq:dispersions} acquires a time dependence from both the mass term ($a^2 m^2$), provided that $m \neq 0$, and the curvature term, provided that $\xi \neq 1/6$.  We assume that if $\xi \geq 1/6$ the effective mass-squared will always be positive, and therefore $\omega_k^2 > 0$ for all $k$.  A value of $\xi$ smaller than $1/6$  implies that some Fourier modes will experience a tachyonic instability ($\omega_k^2 < 0$), which plays an important role in particle production.  

For isotropic field configurations, we can parametrize solutions of the wave equation \eref{eq:RS_mode_eqn_chi} as 
\begin{align}\label{eq:RS_chi_to_alpha_beta}
	\chi_{k}(\eta) = 
	\frac{\alpha_{k}(\eta)}{\sqrt{2 \omega_k(\eta) }} \, e^{- i \theta_k(\eta) }
	+ \frac{\beta_{k}(\eta)}{\sqrt{2 \omega_k(\eta) }} \, e^{i \theta_k(\eta) }
\end{align}
where $\alpha_k$ and $\beta_k$ are complex mode functions, and the phase is defined by $\partial_{\eta} \theta_k(\eta) = \omega_k(\eta)$.  This parametrization is particularly convenient because the Bunch-Davies initial condition becomes $\alpha_k(\eta) \rightarrow 1$ and $\beta_k(\eta) \rightarrow 0$ as $\eta \rightarrow -\infty$.  The parametrization \eref{eq:RS_chi_to_alpha_beta} allows the second-order wave equation for $\chi_k(\eta)$, \eref{eq:RS_mode_eqn_chi}, to be written as a pair of coupled first-order equations for $\alpha_k(\eta)$ and $\beta_k(\eta)$:
\begin{subequations}\label{eq:RS_mode_eqn_alpha_beta}
\begin{align}
	\partial_{\eta} \alpha_{k}(\eta) & = \frac{1}{2} A_{k}(\eta) \, \omega_{k}(\eta) \beta_{k}(\eta) \, e^{2 i \theta_k(\eta)} \\
	\partial_{\eta} \beta_{k}(\eta) & = \frac{1}{2} A_{k}(\eta) \, \omega_{k}(\eta) \alpha_{k}(\eta) \, e^{-2 i \theta_k(\eta)} 
	\per
\end{align}
\end{subequations}
The coefficient $A_k(\eta)$, which is defined by 
\begin{align}\label{eq:adiabatica}
A_k(\eta) & \equiv \frac{\partial_\eta\omega_k(\eta)}{\omega_k^2(\eta)} 
\com
\end{align}
quantifies the departure from adiabaticity, {\it i.e.} it is large if the dispersion relation changes rapidly.  

The abundance of gravitationally produced particles is determined by integrating \eref{eq:RS_mode_eqn_alpha_beta} with the initial condition $\alpha_k(\eta=-\infty)=1$ and $\beta_k(\eta=-\infty)=0$.  At late times the evolution becomes adiabatic ($A_k(\eta) \ll 1$), and the mode functions, $\alpha_k$ and $\beta_k$, become static.  The physical number density of created particles is then given by \eref{eq:n_Bogo}.

%==================================
% Minimally coupled scalar field
\subsubsection{Minimally coupled scalar field}
%==================================

For the minimally coupled scalar field ($\xi = 0$) the dispersion relation \eref{eq:dispersion} and adiabaticity parameter \eref{eq:adiabatica} can be written as 
\begin{subequations}
\begin{align}
\omega^2_k(\eta) & = k^2 + a^2 \left( m^2 - 2 H^2 \right) + a \partial_\eta H \label{eq:MS_omega}\\ 
A_k(\eta) & = \dfrac{\left( m^2 - 2 H^2 \right) a^3 H - \dfrac{3}{2} a^2 H \partial_\eta H + \dfrac{1}{2} a \partial_\eta^2 H}{\left[ k^2 + a^2 \left( m^2 - 2 H^2 \right) + a \partial_\eta H \right]^{3/2}} \label{eq:MS_A} \com
\end{align}
\end{subequations}
where the factor in the denominator of \eref{eq:MS_A} is just $\omega_k^3(\eta)$.  Efficient particle production occurs when $\omega_k^2$ passes through zero and $A_k$ diverges; see \eref{eq:RS_mode_eqn_alpha_beta}.    During inflation we can neglect the term $a \partial_\eta H$.  For $m^2 \geq 2 H^2$ there is no time at which the adiabaticity parameter diverges.  On the other hand, for $m^2 < 2 H^2$, the frequency passes through zero at a time $\eta_k$ such that $k = \sqrt{2} \, a(\eta_k) H(\eta_k) + O(m/H)$.  Since $(aH)^{-1}$ is the comoving Hubble radius, the modes with $|\kvec| = k$ experience their largest departure from adiabaticity at the time of horizon crossing ($k \approx aH$).  Consequently, one expects efficient particle production for light scalar fields, $m^2 \ll 2 H^2$, and little particle production for heavy fields, $m^2 > 2 H^2$.  

\begin{figure}[t]
\begin{center}
\includegraphics[width=0.75\textwidth]{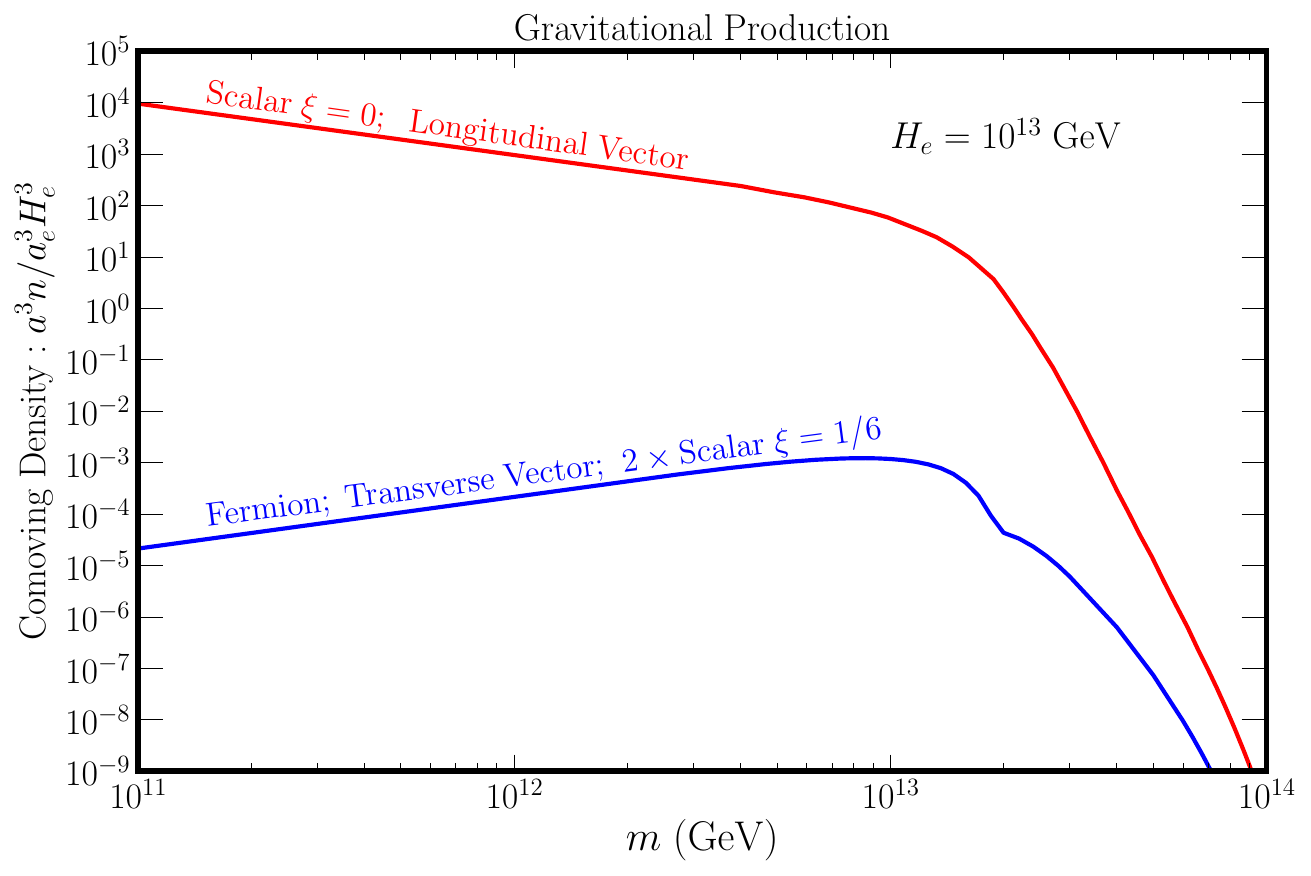}
\caption{\label{fig:n_grav}
The comoving number density of WIMPzilla dark matter produced though its gravitational interaction during or at the end of inflation.}
\end{center}
\end{figure}

The authors of \rref{Kuzmin:1998kk} calculated the relic abundance of gravitationally produced, minimally coupled, scalar dark matter.  We reproduce their result in \fref{fig:n_grav} where we have scaled their calculation\footnote{Figure 2 of \rref{Kuzmin:1998kk} shows $\rho/(\rho_c m_{13}^2)$ where $\rho_c(a) = 3 \Mpl^2 H^2(a)$ is the cosmological critical density and $m_{13}$ is the inflaton mass in units of $10^{13} \GeV$. (The inflaton mass is approximately $2 H_e$.)  This ratio is static during the matter dominated phase of reheating.  We evaluate $(a^3 n / a_e^3 H_e^3) = (3 \Mpl^2 m_{13}^2 / H_e^2)^{-1} (m/H_e)^{-1} (\rho/\rho_c m_{13}^2)$ where $(3 \Mpl^2 m_{13}^2 / H_e^2)^{-1} \simeq (7.1 \times 10^{11})$, which is static at all times after reheating.} to show the comoving number density $a^3 n$ normalized to $a_e^3 H_e^3$.  For $m/H_e < 1$, the scalar field amplitude is fixed to roughly $\langle \phi^2 \rangle \sim H_e^2$ until $H$ drops below $m$ at time $\eta_\ast$, and then the field begins to oscillate about the minimum of its potential, behaving like nonrelativistic matter.  At this time the physical number density is roughly $n = \rho/m \sim m H_e^2$, and the comoving number density is larger by a factor of $(a_\ast/a_e)^3 = H_e^2/m^2$, which explains the scaling $a^3 n / a_e^3 H_e^3 \sim (H_e/m)$.  For $m/H_e > 1$ the gravitational particle production is exponentially suppressed, and $\langle \phi^2 \rangle \ll H_e^2$.  

Generalizing to $\xi \neq 0$, we expect the results to be qualitatively unchanged for any $\xi < 1/6$ (including negative $\xi$), because the dispersion relation \eref{eq:dispersion} admits a tachyonic phase where $\omega_k^2 < 0$ as long as $m^2 < 2(1-6\xi)H^2$.  

%=============
%=============

%==================================
% Conformally coupled scalar field
\subsubsection{Conformally coupled scalar field}
%==================================

For the conformally coupled scalar field ($\xi=1/6$) the dispersion relation \eref{eq:dispersion} and adiabaticity parameter \eref{eq:adiabatica} can be written as 
\begin{align}
\omega_k^2(\eta) & = k^2 + a^2 m^2 \com \quad \textrm{and} \quad
A_{k}(\eta) = \frac{a^2 m^2}{\left[ k^2 + a^2 m^2 \right]^{3/2}} \ (aH) \per
\end{align}
Unlike the case of the minimally coupled scalar field, $A_k$ does not diverge at the time of horizon crossing because $\omega_k^2 > 0$ at all times (we assume $m^2 > 0$).  Instead, $A_k \sim (aH)$ is maximized near the end of inflation, since the end of inflation is defined as the time when the comoving Hubble radius $(aH)^{-1}$ stops decreasing and starts increasing.  In fact, for nonrelativistic modes ($k/a \ll m$) we have $A_k \sim (aH)/a$, which peaks just before the end of inflation, and for relativistic modes ($k/a \gg m$) we have $A_k \sim a^2(aH)$, which peaks just after the end of inflation.  Consequently, the gravitational production of a conformally coupled scalar field primarily occurs at the end of inflation, rather than at the time of horizon crossing.  Since there is no divergence in $A_k(\eta)$, the abundance of gravitationally produced, conformally coupled scalars is expected to be smaller than minimally coupled scalars.  

In \fref{fig:n_grav} we do not show explicitly the result for gravitational production of a conformally coupled scalar field.  However, we will see in the next subsection that the gravitational production of fermions has the same qualitative features as gravitational production of conformally coupled scalars.  Therefore, the comoving number density of gravitationally produced, conformally coupled scalars is well represented in \fref{fig:n_grav} by the curve labeled ``Fermion.''  In the limit $m/H_e \to 0$ the theory enjoys a conformal symmetry, and there is no gravitational particle production \cite{Chung:1998zb}.  In the low-mass regime, $m/H_e \ll 1$, we can understand the scaling with $m$ as approximately $m/H_e$ as follows \cite{Kuzmin:1998kk}:  Since the largest departure from adiabaticity occurs after the end of inflation (see above), the FRW scale factor evolves as $a \propto H^{-\alpha}$ with $\alpha = 2/3$ for a matter dominated universe.  The density of gravitationally produced particles is $n \sim m^3 (a/a_\ast)^{-3}$ where $a_\ast$ is the value of the scale factor when $H$ drops below $m$ and particle production stops.  It follows that $a^3 n / (a_e^3 H_e^3) \sim (m/H_e)^{3-3\alpha}$, which is $(m/H_e)^1$ for $\alpha = 2/3$.  

We expect the results to be qualitatively similar for $\xi \gtrsim 1/6$, because $\omega_k^2(\eta) > 0$ is positive at all times.  However, for larger $\xi$ oscillations of the Ricci scalar during reheating can drive a parametric resonance, which leads to an additional source of particle production \cite{Bassett:1997az,Markkanen:2015xuw}.

%==================================
% Gravitational WIMPzilla Production Fermions
\subsection{Fermions}\label{sec:Gravitational_Fermions}
%==================================

In this section we discuss gravitational particle production for spin-1/2 fermions.  Although we are primarily interested in Majorana fermions, the calculation is more transparent in the case of Dirac fermions.  Here we briefly review the calculation for Dirac fermions following \rref{Chung:2011ck}.  Since particle and antiparticles are produced in equal abundance, due to the universal nature of the gravitational interaction, the Dirac and Majorana calculations differ only by a factor of 2.  

Consider a free Dirac fermion field $\psi(x)$ with minimal gravitational interaction.  The theory is specified by the action
\begin{align}
	S[\psi(x),\overline{\psi}(x)] = \int \! \ud^4 x \, \sqrt{-g} \ \overline{\psi}(x) \bigl( i \gamma^{a} \nabla_{e_{a}} - m \bigr) \psi(x) \per
\end{align}
In an FRW spacetime, the action can be written as 
\begin{align}\label{eq:action_Psi}
	S[\psi(\eta,\xvec),\overline{\psi}(\eta,\xvec)] = \int_{-\infty}^{\infty} \! \ud \eta \int \! \ud^3 \xvec \ \overline{\psi}(\eta, \xvec) \left[ i \gamma^{\mu} \partial_{\mu} - a(\eta) \, m \right] \psi(\eta,\xvec)
\end{align}
after performing a Weyl transformation which absorbs a factor of $a^{3/2}(\eta)$ into the field $\psi$.  The Dirac equation is written as 
\begin{align}\label{eq:Dirac_eqn}
	\left[ i \gamma^{\mu} \partial_{\mu} - a(\eta) \, m \right] \psi(\eta,\xvec) & = 0  \per
\end{align}
This is identical to the Dirac equation in flat (Minkowski) space up to the replacement $m \to a(\eta) \, m$.  This result should not be surprising:  FRW and Minkowski space are conformally equivalent, and the fermion mass $m$ is the only source of conformal symmetry breaking.  Consequently, the spectrum of gravitationally produced particles must vanish as $m \to 0$, similar to the case of the conformally coupled scalar field.  

The field operator $\psi(\eta,\xvec)$ can be decomposed into the mode functions $U_{r,\kvec}(\eta)$ and $V_{r,\kvec}(\eta)$, which are labeled by the wave vector $\kvec$ and the helicity quantum number $r = \pm 1$.  The decomposition is written as 
\begin{align}
	\psi(\eta,\xvec) = \sum_{r = \pm 1} \int \! \! \frac{\ud^3 \kvec}{(2\pi)^3} \, \left( \hat{a}_{\kvec,r} \, U_{r,\kvec}(\eta) e^{i \kvec \cdot \xvec} + \hat{b}_{\kvec,r}^{\dagger} \, V_{r,\kvec}(\eta) e^{-i \kvec \cdot \xvec} \right) 
\end{align}
where $\hat{a}_{\kvec,r}$ and $\hat{b}_{\kvec,r}$ are the annihilation operators.  If we further write the mode functions as 
\begin{align}
	U_{r,\kvec}(\eta) = \begin{pmatrix} u_{A,k}(\eta) \, h_{\kvec,r} \\ r \, u_{B,k}(\eta) \, h_{\kvec,r} \end{pmatrix} 
	\qquad \text{and} \qquad
	V_{r,\kvec}(\eta) = \begin{pmatrix} -u^{\ast}_{B,k}(\eta) \, h_{-\kvec,r} \\ r \, u^{\ast}_{A,k}(\eta)\, h_{-\kvec,r} \end{pmatrix} \, e^{- i r \phi}
\end{align}
where $h_{\kvec,r}$ is an eigenfunction of the helicity operator with eigenvalue $r$ and $\phi$ is the azimuthal angle, then the Dirac equation \eref{eq:Dirac_eqn} becomes 
\begin{align}\label{eq:DF_mode_eqn_uA_uB}
	i \partial_{\eta} \begin{pmatrix} u_{A,k}(\eta) \\ u_{B,k}(\eta) \end{pmatrix}
	= \begin{pmatrix}
	m \, a(\eta) & k \\ k & - m \, a(\eta) 
	\end{pmatrix}
	\begin{pmatrix} u_{A,k}(\eta) \\ u_{B,k}(\eta) \end{pmatrix}
	\per
\end{align}
This equation should be solved along with the Bunch-Davies initial condition
\begin{align}\label{eq:DF_BD_uA_uB}
	\begin{pmatrix} u_{A,k}(\eta) \\ u_{B,k}(\eta) \end{pmatrix}
	\xrightarrow{\eta \to - \infty}
	\begin{pmatrix} u_{A,k}^{\BD}(\eta) \\ u_{B,k}^{\BD}(\eta) \end{pmatrix}
	\equiv 
	\frac{1}{\sqrt{2\omega_{k}(\eta)}}
	\begin{pmatrix}
	\sqrt{\omega_{k}(\eta) + m \, a(\eta)} \\ \sqrt{\omega_{k}(\eta) - m \, a(\eta)} 
	\end{pmatrix}
	e^{- i \theta_{k}(\eta)}
\end{align}
where the dispersion relation, $\omega^2_{k}(\eta) = k^2 + m^2 a^2(\eta)$, is the same one that we encountered for the conformally coupled scalar field, and $\partial_\eta \theta_k = \omega_k$ as before.  
It is convenient to introduce the ansatz 
\begin{align}\label{eq:DF_Ansatz_alpha_beta}
	\begin{pmatrix} u_{A,k}(\eta) \\ u_{B,k}(\eta) \end{pmatrix}
	= 
	\alpha_{k}(\eta) \begin{pmatrix} u_{A,k}^{\BD}(\eta) \\ u_{B,k}^{\BD}(\eta) \end{pmatrix}
	+ \beta_{k}(\eta) \begin{pmatrix} -u_{B,k}^{\BD\ast}(\eta)\\ u_{A,k}^{\BD\ast}(\eta)
 \end{pmatrix}
\end{align}
since the Bunch-Davies initial condition simply becomes $\alpha_k \to 1$ and $\beta_k \to 0$.  
In terms of the mode functions $\alpha_k(\eta)$ and $\beta_k(\eta)$, the mode equations become 
\begin{subequations}\label{eq:DF_mode_eqn_alpha_beta}
\begin{align}
	\partial_{\eta} \alpha_{k}(\eta) & = - \frac{1}{2} A_{k}(\eta) \, \omega_{k}(\eta) \beta_{k}(\eta) \, e^{2i \theta_{k}(\eta)} \\
	\partial_{\eta} \beta_{k}(\eta) & = + \frac{1}{2} A_{k}(\eta) \, \omega_{k}(\eta) \alpha_{k}(\eta) \, e^{-2i \theta_{k}(\eta)}
\end{align}
\end{subequations}
where $A_k(\eta) \equiv mk \partial_\eta a / \omega_k^3(\eta)$.  
Note the strong resemblance with the mode equation for the conformally coupled scalar field that appears in \eref{eq:RS_mode_eqn_alpha_beta}.  

We solve \eref{eq:DF_mode_eqn_alpha_beta} numerically, extract the late-time behavior of $\beta_k(\eta)$, and calculate the comoving number density of gravitationally produced particles using \eref{eq:n_Bogo}.  For a Dirac fermion one would take $g = 4$ in \eref{eq:n_Bogo}, which counts two spin states and two particle/anti-particle states, but we take $g = 2$ to count only the two spin states of the Majorana fermion.  

Figure \ref{fig:n_grav} also shows the predicted comoving number density of gravitationally produced  fermion WIMPzillas.  For Dirac fermions, the abundance would be larger by a factor of $2$ and for conformally coupled scalars, the abundance would be qualitatively similar and smaller by a factor of approximately $2$.  In the small-mass regime, $m/H_e \ll 1$, the theory enjoys an approximate conformal symmetry, and the abundance is suppressed.

%==================================
% Gravitational WIMPzilla Production Vectors
\subsection{Vectors}\label{sec:Gravitational_Vectors}
%==================================

Finally we review the gravitational production of spin-1 vector particles; additional details can be found in Refs.~\cite{Dimopoulos:2006ms,Graham:2015rva}.  Consider a neutral vector field $A_\mu(x)$ with a minimal gravitational interaction.\footnote{More generally, we could introduce a nonminimal coupling such as $\xi R A_\mu A^\mu$.  However, this operator does not respect the gauge invariance under which $A_\mu \to A_\mu + \partial_\mu \chi$, and thus one expects $\xi \sim m^2/\Lambda^2$.  During inflation this term contributes to the vector mass on the order of $m^2H^2/\Lambda^2$, but the validity of the EFT requires $H^2 / \Lambda^2 \ll 1$, so this term is negligible compared to the usual mass term.}  The action for this theory is written as 
\begin{align}
	S = \int \! \ud^4 x \, \sqrt{-g} \left[ - \frac{1}{4} g^{\mu \alpha} g^{\nu \beta} F_{\mu \nu} F_{\alpha \beta} - \frac{1}{2} m^2 g^{\mu \nu} A_{\mu} A_{\nu} \right] 
	\per
\end{align}
In an FRW spacetime the action becomes 
\begin{align}
	S = \int_{-\infty}^{\infty} \! \ud \eta \int \! \ud^3 \xvec \, \frac{1}{2} \left[ \left| \partial_\eta \Avec - a^2 {\bm \nabla} A_0 \right|^2 - \left| {\bm \nabla} \times \Avec \right|^2 + a^4 m^2 A_0^2 - a^2 m^2 | \Avec |^2 \right] 
	\per 
\end{align}
The field $A_0$ does not have a kinetic term, and therefore we can integrate it out exactly.  This is most easily done by first moving to Fourier space where we have 
\begin{align}
	S = \int_{-\infty}^{\infty} \! \ud \eta \int \! \! \frac{\ud^3 \kvec}{(2\pi)^3} \, & \left\{ 
	\frac{1}{2} (k^2 + a^2 m^2) \left| a A_0 - \frac{ i {\bm k} \cdot \partial_\eta \Avec}{k^2 + a^2 m^2} \right|^2 \right. \nn  & \quad \left.
+\frac{1}{2} \left( | \partial_\eta \Avec |^2 - \frac{|{\bm k} \cdot \partial_\eta \Avec|^2}{k^2 + a^2 m^2}- | {\bm k} \times \Avec |^2 - a^2 m^2 |\Avec|^2 \right) 
	\right\}
	\per 
\end{align}
Now the integral over $A_0$ is Gaussian, and we can integrate it out trivially.  Next we write $\Avec = \Avec_T + \Avec_L$ where $\Avec_T$ represents the two transverse polarization modes (${\bm k} \cdot \Avec_T = 0$ and ${\bm k} \times \Avec_T = \pm k |\Avec_T|$) and $\Avec_L$ represents the single longitudinal polarization mode (${\bm k} \cdot \Avec_L = k A_L$ and ${\bm k} \times \Avec_L = 0$).  
Then, the action breaks up into $S = S_T + S_L$ where 
\begin{subequations}
\begin{align}
S_T & = \int_{-\infty}^{\infty} \! \ud \eta \int \! \! \frac{\ud^3 \kvec}{(2\pi)^3} \, \frac{1}{2} \left[ \left| \partial_\eta \Avec_T \right| ^2 - \left( k^2 + a^2 m^2 \right) |\Avec_T|^2 \right] \\ 
S_L & = \int_{-\infty}^{\infty} \! \ud \eta \int \! \! \frac{\ud^3 \kvec}{(2\pi)^3} \,\frac{1}{2} \left[ \frac{a^2 m^2}{k^2 + a^2 m^2} \left( \partial_\eta A_L \right)^2 - a^2 m^2 A_L^2 \right] \per 
\end{align}
\end{subequations}

The two transversely polarized modes are canonically normalized, and we can immediately read off the dispersion relation, which is just $\omega_k^2(\eta) = k^2 + a^2(\eta) m^2$.  This is the same dispersion relation that we encountered when studying the conformally coupled scalar field; see \eref{eq:dispersion} with $\xi = 1/6$.  Thus, the abundance of gravitationally produced, \textit{transversely polarized} spin-1 particles is simply double the abundance of conformally coupled scalar particles of the same mass.  This is represented by the blue curve in \fref{fig:n_grav}.  As before, the suppression at small mass is understood, because an enhanced conformal symmetry arises when $m/H_e \to 0$.  

The longitudinally polarized mode is more complicated.  In order to have a canonical kinetic term we define a new field $\phi_L$ in terms of $A_L$ by $A_L = (am)^{-1}\sqrt{k^2 + a^2 m^2} \, \phi_L$.  
Then $S_L$ becomes, 
\begin{align}
S_L &  = \int_{-\infty}^{\infty} \! \ud \eta \int \! \! \frac{\ud^3 \kvec}{(2\pi)^3} \, \frac{1}{2} 
\biggl\{ \left(\partial_\eta\phi_L\right)^2 - 2aH \partial_\eta \phi_L \left(1 - \frac{a^2 m^2}{k^2 + a^2 m^2}\right)\phi_L \nn & \qquad \qquad
+ a^2 H^2 \left(1-\frac{a^2 m^2}{k^2 + a^2 m^2}\right)^2 \phi_L^2 - \left( k^2 + a^2 m^2 \right) \phi_L^2 \biggr\} \per 
\end{align}
Then integrating by parts and dropping the total derivative gives  
\begin{align}
S_L & = \int_{-\infty}^{\infty} \! \ud \eta \int \! \! \frac{\ud^3 \kvec}{(2\pi)^3} \, \frac{1}{2} \left\{ \left( \partial_\eta \phi_L \right)^2 - \left[ k^2 + a^2  m^2 - a^2\frac{R}{6}  \right. \right. \nn & \qquad \qquad \left. \left. + a^2\frac{R}{6} \left(1-\frac{k^2}{k^2+a^2m^2}\right)
+3a^2H^2\left(\frac{k^2}{k^2+a^2m^2}\right)\left(1-\frac{k^2}{k^2+a^2m^2}\right) 
\right] \phi_L^2 \right\} \per 
\end{align}
Since the kinetic term is now canonically normalized, the dispersion relation is simply equal to the expression in square brackets.  In general this expression cannot be matched to the dispersion relation for a scalar field \eref{eq:dispersion}, even with a judicious choice of the nonminimal coupling $\xi$.  However, a matching can be performed in limiting regimes.  For the nonrelativistic modes, $k/a \ll m$, the dispersion relation becomes $\omega_k^2 \approx k^2 + a^2 m^2$, which matches the conformally coupled scalar field model.  For the relativistic modes, $k/a \gg m$, the dispersion relation becomes $\omega_k^2 \approx k^2 + a^2 m^2 - a^2 R/6$, which matches onto the minimally coupled scalar field model.\footnote{As expected from the Goldstone boson equivalence theorem, the relativistic limit of a massive vector field behaves as a massless vector plus a minimally coupled scalar field, which corresponds to the eaten Goldstone boson.}  Since we have seen in \fref{fig:n_grav} that gravitational particle production is much more efficient for minimally coupled scalars, we expect that the longitudinal polarization modes will be efficiently produced in the regime $m \ll H$ with the largest departure from adiabaticity occurring when a mode exits the horizon, $m \ll k/a \sim H$.  Then the comoving density of longitudinally polarized vectors is well approximated by the red curve in \fref{fig:n_grav}, and this population dominates over the transversely polarized vectors.

%==================================
% Summary of gravitational production
\subsection{Summary of gravitational production of WIMPzillas}\label{sec:Gravitational_Summary}
%==================================

The results are summarized in \fref{fig:n_grav} where we show the comoving WIMPzilla number density normalized to the comoving Hubble volume at the end of inflation, $1/(a_e^3H_e^3)$.  The comoving density $a^3 n$ is static and the physical density $n$ redshifts like $a^{-3}$, as we expect for a dark matter candidate.  One can derive the corresponding relic abundance $\Omega h^2$ using the formulas in \sref{sec:ThermalDM}.  The numerical results are well approximated by the following empirical formulas, which show the comoving number density well after inflation when $t \gg m$
\label{eq:n_grav}
\begin{align}
%\text{Scalar ($\xi = 0$) \& longitudinal vector:} 
\left. \begin{array}{r}
\mathrm{Scalar}\ (\xi=0) \\
\mathrm{Longitudinal\ Vector}
\end{array} \right \}
& \qquad 
\dfrac{a^3 n}{a_e^3 H_e^3} \simeq \qquad \begin{cases}
96  \dfrac{H_e}{m} & \ \phantom{00}\dfrac{m}{H_e}<1 \\[4pt]
0.76  \dfrac{H_e}{m} \, e^{-2m/H_e} & \  \phantom{00}\dfrac{m}{H_e} >1 \nn
\end{cases} \\[8pt]
%--
%\text{Fermion,}\ 2\times \text{Scalar}\ (\xi=1/6),\ \text{\& transverse\ vector:} 
\left.\begin{array}{r}
\mathrm{Fermion} \\
2\times\ \mathrm{Scalar}\ (\xi=1/6) \\
\mathrm{Transverse\ Vector}
\end{array} \right\}
& \qquad 
\dfrac{a^3 n}{a_e^3 H_e^3} \simeq \qquad \begin{cases}
0.0021  \dfrac{m}{H_e} & \ \dfrac{m}{H_e}<1 \\[4pt]
0.0080  \dfrac{H_e}{m} \, e^{-2m/H_e} & \ \dfrac{m}{H_e} >1 \per \\
\end{cases}
\end{align}

For these estimates, we have assumed that the WIMPzilla is a self-conjugate particle, but if there are multiple species of degenerate WIMPzilla particles and antiparticles, the redundancy is taken into account by a trivial rescaling of $g$ in \eref{eq:n_Bogo}.

%==================================
% Thermal WIMPzilla Production
%==================================
\section{Thermal Production of WIMPzillas}\label{sec:Thermal}

We now turn to the main purpose of this paper: to find the parameters where thermal production of supermassive particles will dominate gravitational production.  In the last section we reviewed the calculation of gravitational production of WIMPzillas.  In this section we calculate thermal production of WIMPzillas.  We suppose that the WIMPzilla interacts with the SM particles through the Higgs portal, and we calculate the number density of WIMPzilla particles that are produced from Higgs annihilations in the early universe.  

The interactions in \eref{eq:interactions} allow WIMPzilla pairs to be produced from the annihilation of Higgs-boson pairs.  At the temperatures of interest ($T \gg 100 \GeV$) the electroweak symmetry is unbroken, and the Higgs field factor $\Phi^\dagger\Phi$ represents two states.  The two WIMPzilla production channels are 
\begin{align}\label{eq:two_channels}
	\Phi^0 \bar{\Phi}^0 \longrightarrow XX
	\qquad \text{and} \qquad 
	\Phi^+ \Phi^- \longrightarrow XX
	\com
\end{align}
where we use $X$ to denote the WIMPzilla whose identity is yet unspecified.  The physical number density of WIMPzilla particles satisfies the kinetic equation for self-conjugate particles
\begin{align}\label{eq:kinetic_eqn}
\dot{n} + 3 H n = - \langle \sigma v \rangle \left( n^2 - \bar{n}^2 \right) \com
\end{align}
where $\langle \sigma v \rangle$ is the time-dependent (and hence, temperature-dependent) thermally averaged WIMPzilla annihilation cross section.  The equilibrium density of WIMPzilla particles with mass $m > T$ is denoted by $\bar{n}$, and it takes the value (for Maxwell-Boltzmann statistics)
\begin{align}\label{eq:neq_def}
	\bar{n}(t)
	= g \int \! \! \frac{\ud^3 {\bm p}}{(2\pi)^3} \, e^{-E/T} 
	= g \frac{m^2 T}{2\pi^2} \, K_2(m/T) \per
\end{align}
Here $K_n(x)$ is a modified Bessel function of the second kind of order $n$.  

In the parameter regime of interest, the coupling of the WIMPzilla to the plasma is so weak that the WIMPzilla abundance does not reach the thermal abundance, $n \ll \bar{n}$.  In this regime, the right side of the kinetic equation \eref{eq:kinetic_eqn} reduces to a source term, 
\begin{align}\label{eq:source_def}
	\Scal(t) \equiv \langle \sigma v \rangle \, \bar{n}^{2}
	\com 
\end{align}
which accounts for WIMPzilla production via Higgs boson annihilation.  Using the approximation $n \ll \bar{n}$, we integrate \eref{eq:kinetic_eqn} directly to find the comoving number density of thermally produced WIMPzilla particles:
\begin{align}\label{eq:formal_soln_a}
\frac{a^3(t) n(t)}{a_e^3H_e^3}  = \int_{a_e}^{a(t)} \frac{\ud a^{\prime}}{a_e} \, \frac{a^{\prime 2}}{a_e^2} \, \frac{\Scal(a^{\prime})}{H_e^3H(a^{\prime})} 
	\per
\end{align}
Here the time dependence is captured by the monotonically growing scale factor.  

To evaluate \eref{eq:formal_soln_a} we must know $T(a)$ and $H(a)$, which requires us to specify a model of reheating.  We assume that reheating proceeds through the perturbative decay of the inflaton condensate.  Provided that thermalization occurs quickly, it is known \cite{Chung:1998rq} that the plasma temperature scales as $T \sim a^{-3/8}$ during the epoch of reheating, while the universe remains matter dominated ($H \sim a^{-3/2}$).  When the entropy injection from the inflaton decay is completed, the universe is radiation dominated ($H \sim a^{-2}$) and temperature scales as $T \sim a^{-1}$.  Thus, we model the background evolution as 
\begin{align}
	T(a) & = \begin{cases} 
	T_{\rm max} \left( a / a_e \right)^{-3/8} & \mathrm{for}\ a_e \leq a < a_{\RH} \\
	T_{\RH} \left( a / a_{\RH} \right)^{-1} & \mathrm{for}\ a_{\RH} \leq a 
	\end{cases} \\ 
	H(a) & = \begin{cases} 
	H_e \left( a / a_e \right)^{-3/2} & \mathrm{for}\ a_e \leq a < a_{\RH} \\
	H_{\RH} \left( a / a_{\RH} \right)^{-2} & \mathrm{for}\ a_{\RH} \leq a \com
	\end{cases} 
\end{align}
where $a_{\RH}$ is the value of the scale factor at the start of the radiation era, and we can relate \cite{Chung:1998rq}
\begin{align}\label{eq:Tmax_def}
	\left( \frac{a_{\RH}}{a_{e}} \right)^3 = \left( \frac{T_{\rm max}}{T_{\RH}} \right)^{8} = \left( \frac{H_{e}}{H_{\RH}} \right)^{2} 
	\per
\end{align}
Applying the Friedmann equation to the radiation dominated universe at $a = a_{\RH}$ gives 
\begin{align}\label{eq:HRH_def}
	3 H_{\RH}^2 \Mpl^2 = \frac{\pi^2}{30} g_{\ast} T_{\RH}^4 \com
\end{align}
where $g_{\ast}$ is the effective number of relativistic species at temperature $T_{\RH}$.  
We will use $g_{\ast} = 106.75$, and our results are insensitive to $O(1)$ changes in this value which would arise from new physics above the weak scale.  Using these relations, there are only two free parameters: the Hubble parameter at the end of inflation, $H_e$, and the plasma temperature at the beginning of radiation domination, $T_{\RH}$.  Then, \erefs{eq:Tmax_def}{eq:HRH_def} imply 
\begin{align}\label{eq:Tmax}
	T_{\rm max} 
	%= \left( \frac{\pi^2}{90} g_{\ast} \right)^{-1/8} T_{\RH}^{1/2} H_e^{1/4} \Mpl^{1/4} 
	\simeq \bigl( 1.6 \times 10^{12} \GeV \bigr) \left( \frac{g_{\ast}}{106.75} \right)^{-1/8} \left( \frac{T_{\RH}}{10^9 \GeV} \right)^{1/2} \left( \frac{H_e}{10^{13} \GeV} \right)^{1/4} 
	\per 
\end{align}
As we already mentioned in \sref{sec:Gravitational}, we focus on $H_e = 10^{13} \GeV$, but we take $T_{\RH}$ as a free parameter.  

In the following subsections, we consider each of the WIMPzilla models in turn.  We calculate the thermally averaged annihilation cross section $\langle \sigma v \rangle$,  and we evaluate the abundance of thermally produced particles using \eref{eq:formal_soln_a}.  

%----------------------------------------------------------------
% Scalar WIMPzilla
%----------------------------------------------------------------
\subsection{Scalars}\label{sub:scalar_WIMPzilla}

\begin{figure}[t]
\begin{center}
\includegraphics[width=0.75\textwidth]{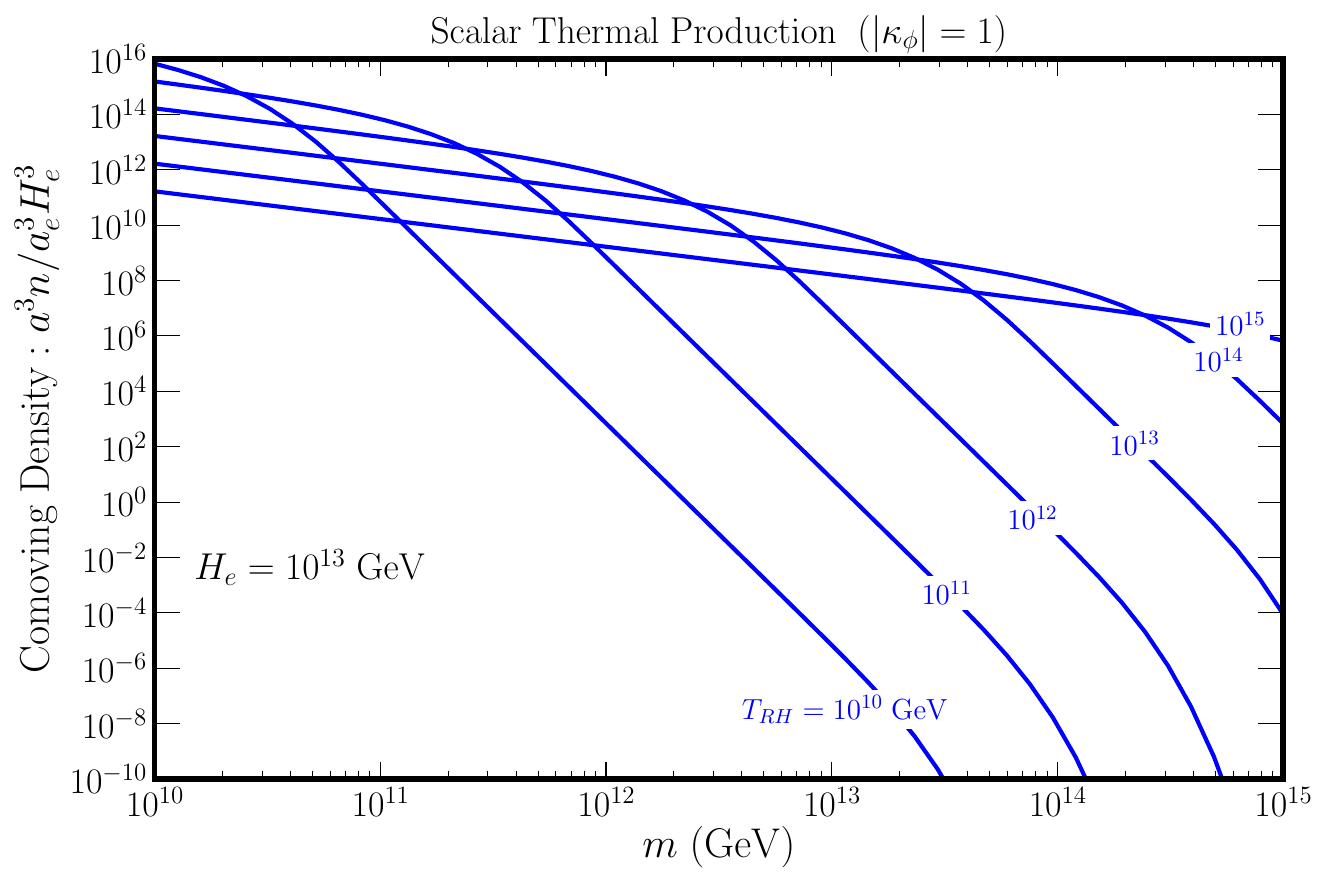}
\caption{\label{fig:scalar_density}
The comoving number density of thermally produced particles in the scalar WIMPzilla model.  We take the Hubble scale at the end of inflation to be $H_e = 10^{13} \GeV$ while varying the WIMPzilla mass $m$ and the reheating temperature $T_{\RH}$.}
\end{center}
\end{figure}

\begin{figure}[t]
\begin{center}
\includegraphics[width=0.49\textwidth]{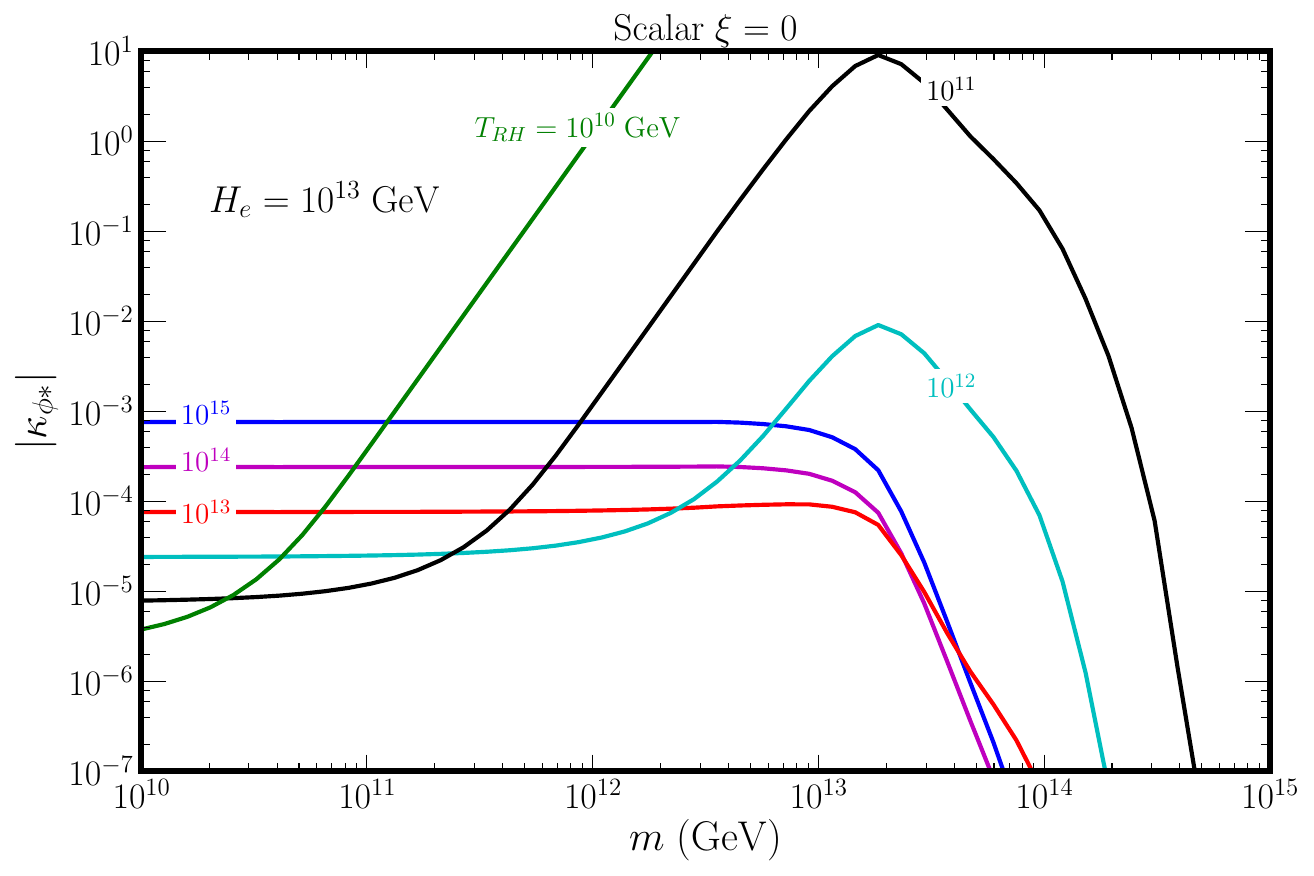} \hfill
\includegraphics[width=0.49\textwidth]{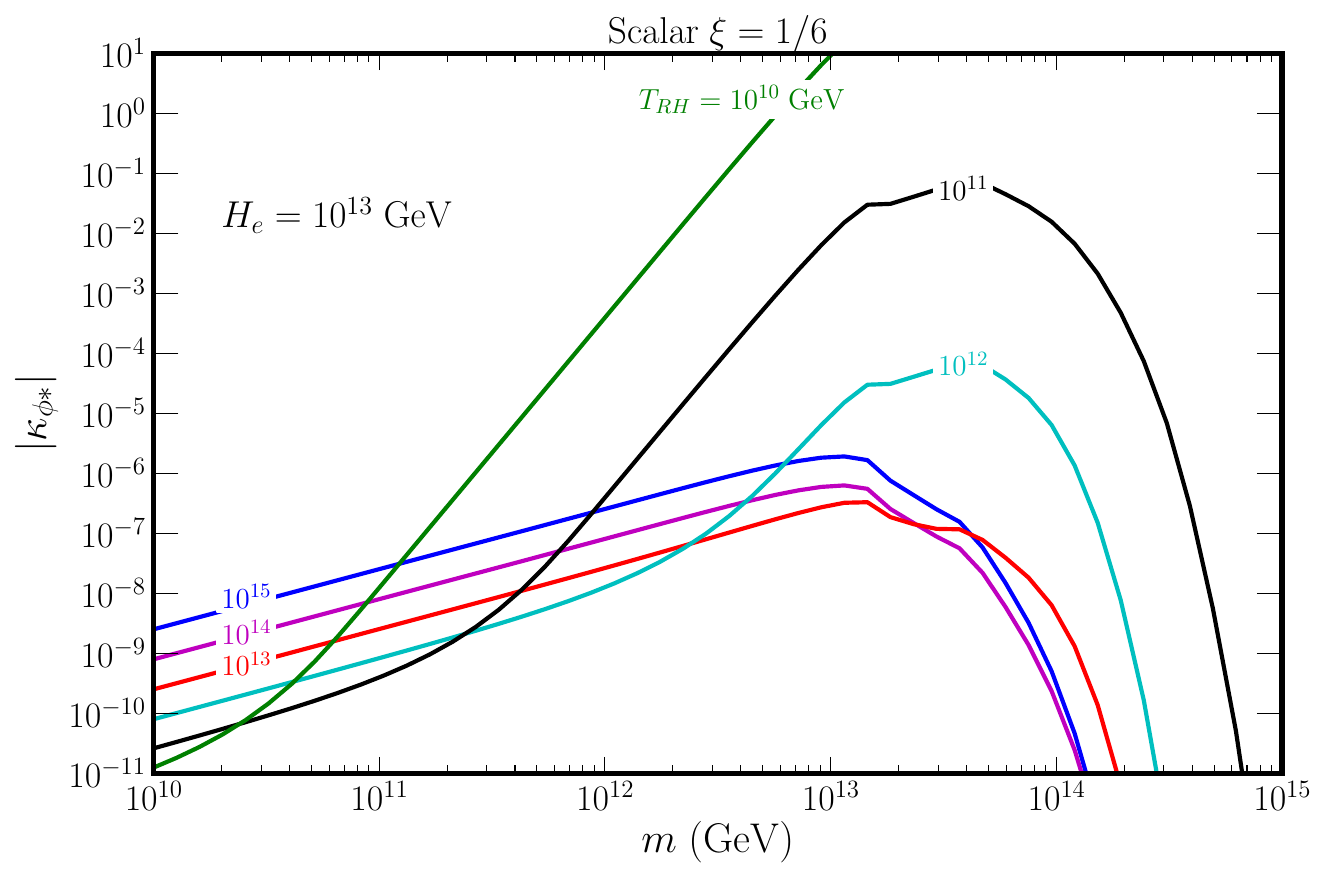} 
\caption{\label{fig:kappa*s}
The values of $\left|\kappa_\phi\right|$ corresponding to equal thermal and gravitational production, denoted as $\left|\kappa_{\phi*}\right|$, for scalar WIMPzilla models for minimally coupled scalars (left panel) and conformally coupled scalars (right panel).}
\end{center}
\end{figure}

If the WIMPzilla is a spin-0 self-conjugate scalar field $\phi(x)$, then the coupling of $\phi$ to the Higgs field, given in \eref{eq:interactions}, is specified by the dimensionless coupling constant $\kappa_\phi$.  It is straightforward to calculate the annihilation cross section; see \aref{app:cross_section}.  Summing over the two channels in \eref{eq:two_channels}, we find the thermally averaged WIMPzilla annihilation cross section to be 
\begin{align}\label{eq:scalar_sigma}
	\langle \sigma v \rangle & = 
	\frac{\left|\kappa_\phi\right|^2}{16 \pi} \, \frac{1}{m^2} \, \frac{K_1^2(m/T)}{K_2^2(m/T)} \per
\end{align}
The source term is then calculated using \erefs{eq:neq_def}{eq:source_def} with $g =1$, and we obtain 
\begin{align}\label{eq:scalar_source}
	\Scal = \frac{\left|\kappa_\phi\right|^2}{64 \pi^5} \, m^2 T^2 \,  K_1^2(m/T) \per
\end{align}
Evaluating the integral in \eref{eq:formal_soln_a} yields the density of thermally produced WIMPzilla particles in terms of special functions.  At late times ($a \gg a_{\RH}$) the source vanishes and the comoving number density of thermally produced WIMPzilla particles becomes static.  Extracting the asymptotic behavior in the small- and large-mass regimes, we find  
\begin{align}\label{eq:scalar_nX}
	\dfrac{a^3 n}{a_e^3H_e^3} \approx \begin{cases}
	\dfrac{105 \, |\kappa_\phi|^2}{64\pi^4} \dfrac{T_{\rm max}^{12}}{H_e^4 m^8} \ e^{-2 m / T_{\rm max}} \, f_0(m/T_{\rm max}) & \textrm{for} \ T_\RH \ll m \\[10pt]
	\dfrac{3 |\kappa_\phi|^2}{2048\pi^3} \dfrac{T_{\rm max}^{12}}{H_e^4 m T_{\RH}^7} & \textrm{for} \ m \ll T_{\RH}
	\end{cases}
\end{align}
where $f_0(x) \equiv 1 + 2x + 2 x^2 + 4 x^3/3 + 2 x^4/3 + 4x^5/15 + 4x^6/45 + 8x^7/315 + 2x^8/315$.    In \fref{fig:scalar_density} we show the comoving number density of thermally produced WIMPzillas.  In \fref{fig:kappa*s} we solve for $\left|\kappa_{\phi*}\right|$, the value of $\left|\kappa_\phi\right|$ that results in an equal population of thermally produced WIMPzillas and gravitationally produced WIMPzillas.  For $\left|\kappa_\phi\right| > \left|\kappa_{\phi*}\right|$ thermal production dominates, while for $\left|\kappa_\phi\right|< \left|\kappa_{\phi*}\right|$ gravitational production dominates. 

%=============
%=============
%=============

%----------------------------------------------------------------
%  Fermion WIMPzilla
\subsection{Fermions}\label{sub:fermion_WIMPzilla}
%----------------------------------------------------------------

%========
If the WIMPzilla is a spin-1/2 fermion $\psi$, then the coupling of the $\psi$ to the Higgs field, given in \eref{eq:interactions}, is specified by the ratio $\kappa_\psi/\Mpl$.  The thermally averaged WIMPzilla annihilation cross section is calculated in \aref{app:cross_section}, and we find 
\begin{align}\label{eq:fermion_sigma}
\langle \sigma v \rangle  = \frac{1}{4\pi} \frac{\left|\kappa_\psi\right|^2}{\Mpl^2} 
	\frac{T^4}{m^4 K_2^2(m/T)} 
	\left[ \frac{3\sqrt{\pi}}{8} G^{30}_{13} \left( \frac{m^2}{T^2}  \left| \begin{array}{l} 5/2 \\ 0, 2, 3  \end{array} \right. \right) \right]  \com
\end{align}
where we have averaged over the $g = 2$ spin states.  Here $G^{mn}_{pq}\left(  z \left| \begin{array}{l} a_1, \ldots, a_p \\ b_1, \ldots, b_q \end{array} \right. \right)$ is the Meijer $G$-function, defined by a line integral in the complex plane
\begin{align}
G^{mn}_{pq} \left( z  \left| \begin{array}{l} a_1,\ldots,a_p \\ b_1,\ldots,b_q  \end{array} \right. \right) = \frac{1}{2\pi i}\int_\gamma\frac
{\prod_{j=1}^m\Gamma(b_j-s)\prod_{j=1}^n\Gamma(1-a_j+s)}{\prod_{j=n+1}^p\Gamma(a_j-s)\prod_{j=m+1}^q\Gamma(1-b_j+s)}x^s ds \com
\end{align}
where $\Gamma(z)$ is the gamma function and $\gamma$ indicates the appropriate contour \cite{meijerG:2013}.  The source term is calculated using \erefs{eq:neq_def}{eq:source_def}, and we obtain 
\begin{align}\label{eq:fermion_source}
	\Scal & = 
	\frac{|\kappa_\psi|^2}{\Mpl^2} 
	\frac{T^6}{4 \pi^5} 
	\left[ \frac{3 \sqrt{\pi}}{8} G^{30}_{13}\left( \frac{m^2}{T^2} \left| \begin{array}{l} 5/2 \\ 0, 2, 3 \end{array} \right. \right) \right] 
	\per
\end{align}
In the limits of asymptotically small and large WIMPzilla mass, these formulas have the following limiting behavior: 
\begin{align}
%---
\langle \sigma v \rangle 
& \approx \dfrac{1}{16\pi} \dfrac{|\kappa_\psi|^2}{\Mpl^2} \times \begin{cases}
\dfrac{3T}{m} & \textrm{for} \ T \ll m \\
1 & \textrm{for} \ m \ll T 
\end{cases}
	\\[6pt]
%---
\Scal 
& \approx 
\dfrac{1}{16\pi} \dfrac{|\kappa_\psi|^2}{\Mpl^2} \times \begin{cases}
T^6\dfrac{3}{2 \pi^3} \dfrac{m^2}{T^2} \, e^{-2m/T} & \textrm{for} \ T \ll m \\[4pt]
T^6\dfrac{4}{\pi^4} & \textrm{for} \ m \ll T \per
\end{cases}
\end{align}
We calculate the comoving number density of thermally produced WIMPzilla particles by evaluating the integral in \eref{eq:formal_soln_a}.  The integral can be expressed in terms of the Meijer G-function in general, and in the asymptotic limits it simplifies to 
\begin{align}\label{eq:fermion_nX}
\frac{a^3 n}{a_e^3H_e^3} \approx 
	\begin{cases}
	\dfrac{945 \, |\kappa_\psi|^2}{128\pi^4} \dfrac{T_{\rm max}^{12}}{H_e^4 \Mpl^2 m^6} e^{-2 m / T_{\rm max}} \, f_{1/2}(m/T_{\rm max}) & \textrm{for} \ T_{\RH} \ll m \\ 
	\dfrac{13|\kappa_\psi|^2}{36\pi^5} \dfrac{T_{\rm max}^{12}}{H_e^4 \Mpl^2 T_{\RH}^6} & \textrm{for} \ m \ll T_{\RH} \com
	\end{cases}
\end{align}
where $f_{1/2}(x) \equiv 1 + 2x + 2 x^2 + 4 x^3/3 + 2 x^4/3 + 4x^5/15 + 4x^6/45 + 16x^7/945$.  

\begin{figure}[t]
\begin{center}
\includegraphics[width=0.49\textwidth]{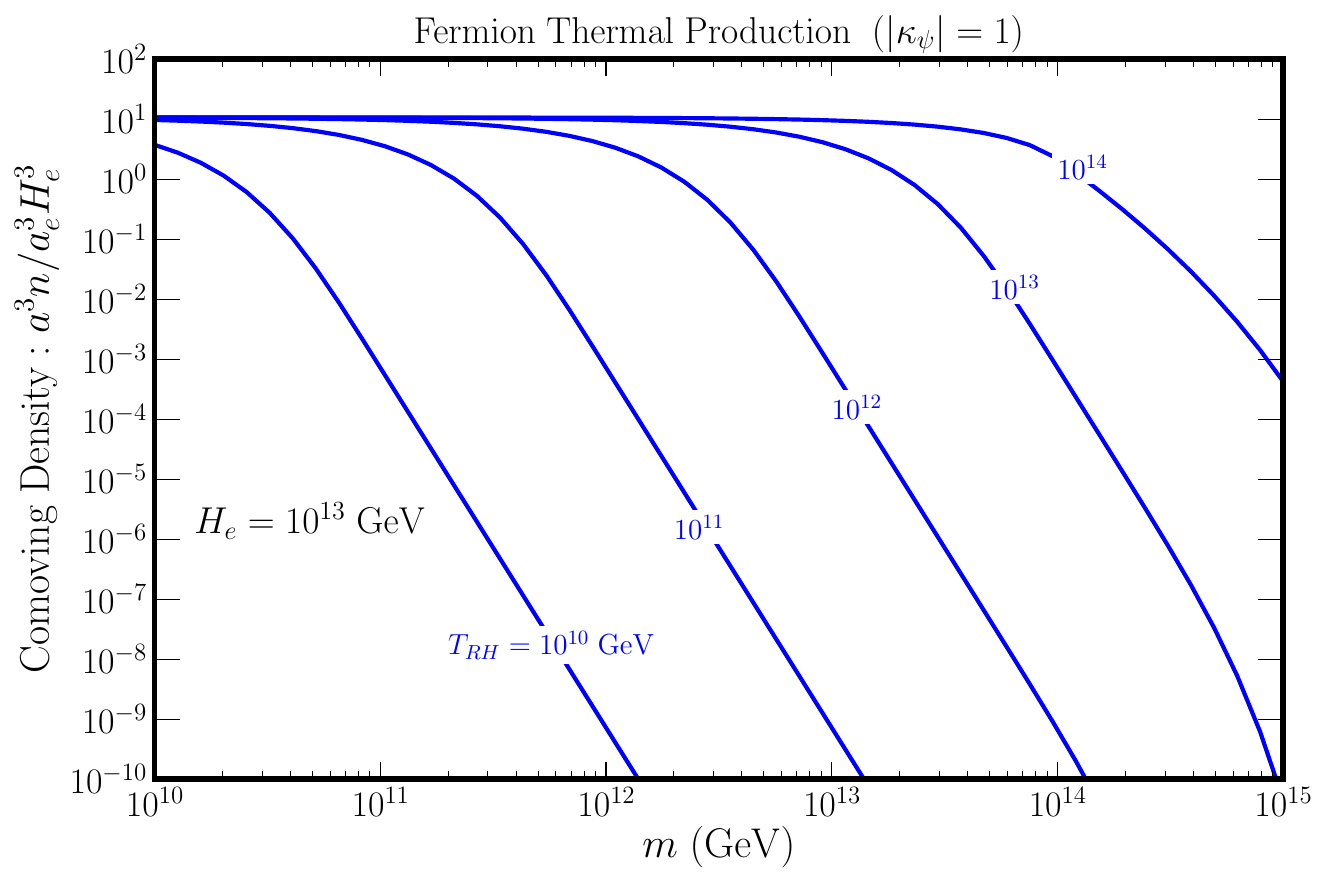} \hfill
\includegraphics[width=0.49\textwidth]{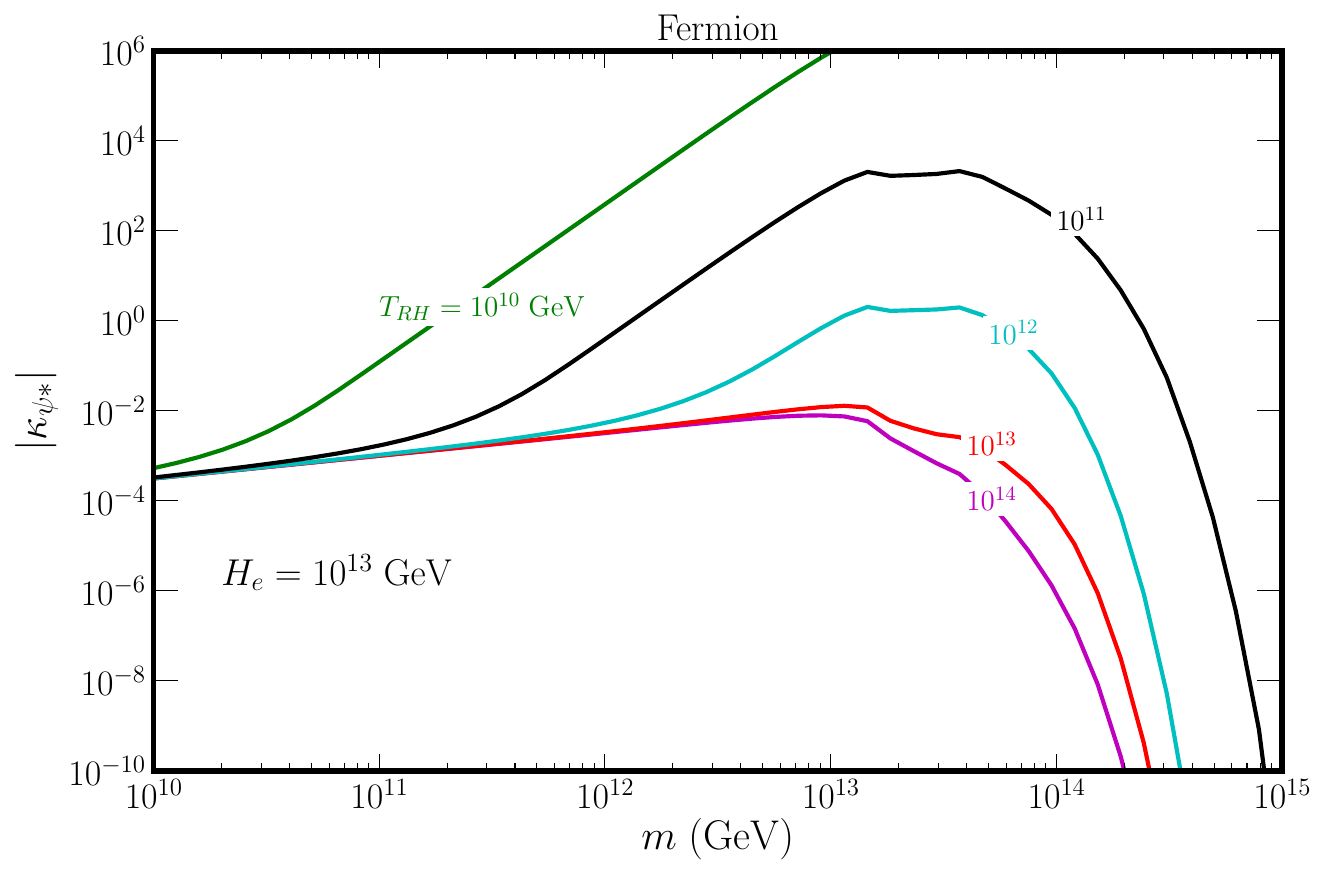}
\caption{\label{fig:fermion_density}
Left panel: The comoving number density of thermally produced particles in the  fermion WIMPzilla model.  We take the Hubble scale at the end of inflation to be $H_e = 10^{13} \GeV$ while varying the WIMPzilla mass $m$ and the reheating temperature $T_{\RH}$. Right panel: The values of $\left|\kappa_\psi\right|$ corresponding to equal thermal and gravitational production, denoted as $\left|\kappa_{\psi*}\right|$, for the fermion WIMPzilla model.}
\end{center}
\end{figure}

In the left panel of \fref{fig:fermion_density} we plot the thermally produced comoving density.  In the right panel we compare the gravitationally produced WIMPzilla abundance to the thermally produced WIMPzilla abundance, and obtain a value of $|\kappa_\psi|$ where the two sources of WIMPzillas will result in equal abundances.   For $\left|\kappa_\psi\right| > \left|\kappa_{\psi*}\right|$, thermal production dominates, while for $\left|\kappa_\psi\right|< \left|\kappa_{\psi*}\right|$, gravitational production dominates.  In the region of parameter space where $|\kappa_{\psi}| \gg 1$, the cutoff of the theory is lowered to $\Mpl / |\kappa_\psi|$.  As we discussed before \eref{eq:kappamax}, the validity of the EFT imposes $|\kappa_\psi| \ll 10^6$, which is satisfied across  the entire parameter space shown in \fref{fig:fermion_density}.  

%=============

%----------------------------------------------------------------
% Vector WIMPzilla
\subsection{Vectors}\label{sub:vector_WIMPzilla}
%----------------------------------------------------------------

\begin{figure}[t]
\begin{center}
\includegraphics[width=0.49\textwidth]{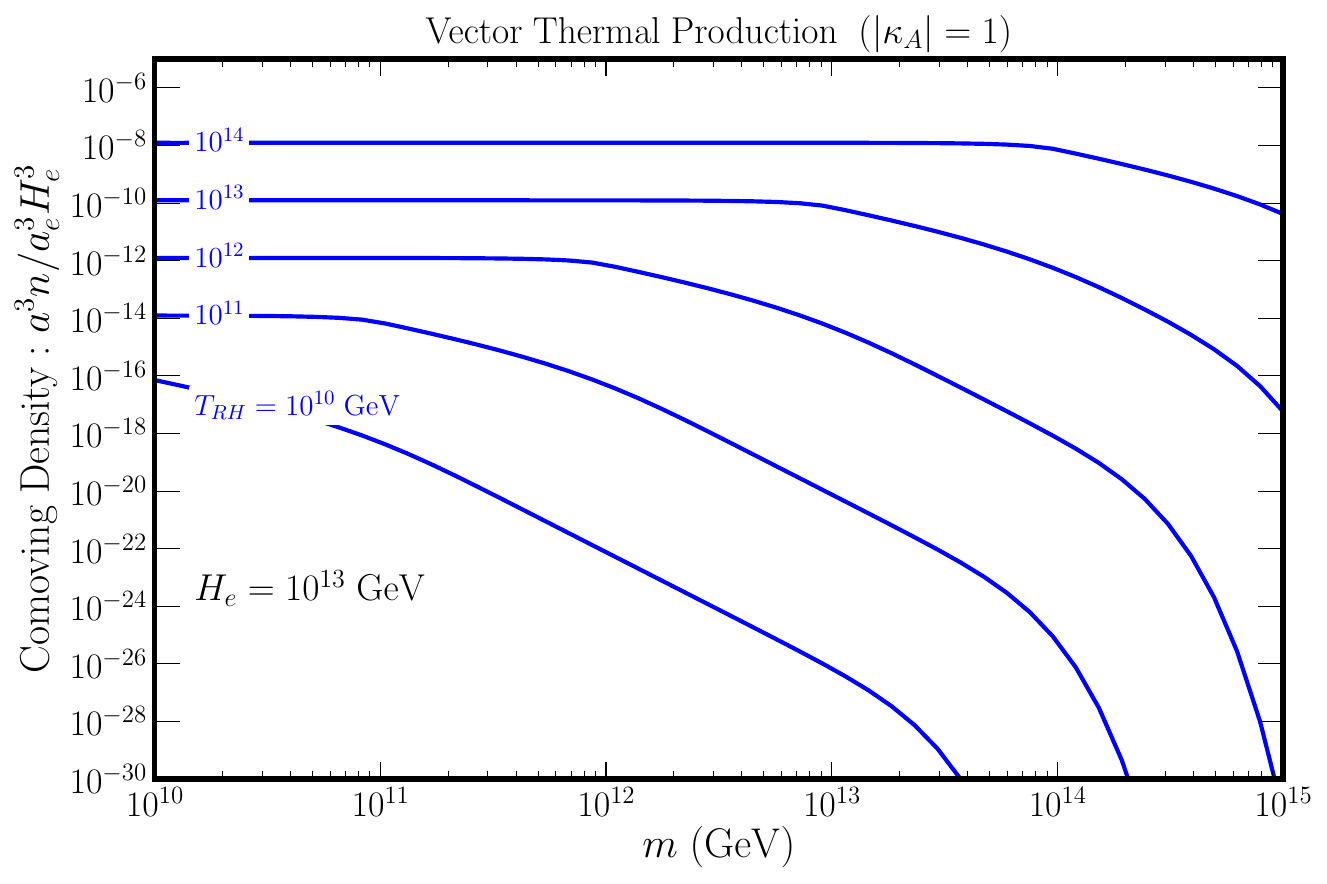} \hfill
\includegraphics[width=0.49\textwidth]{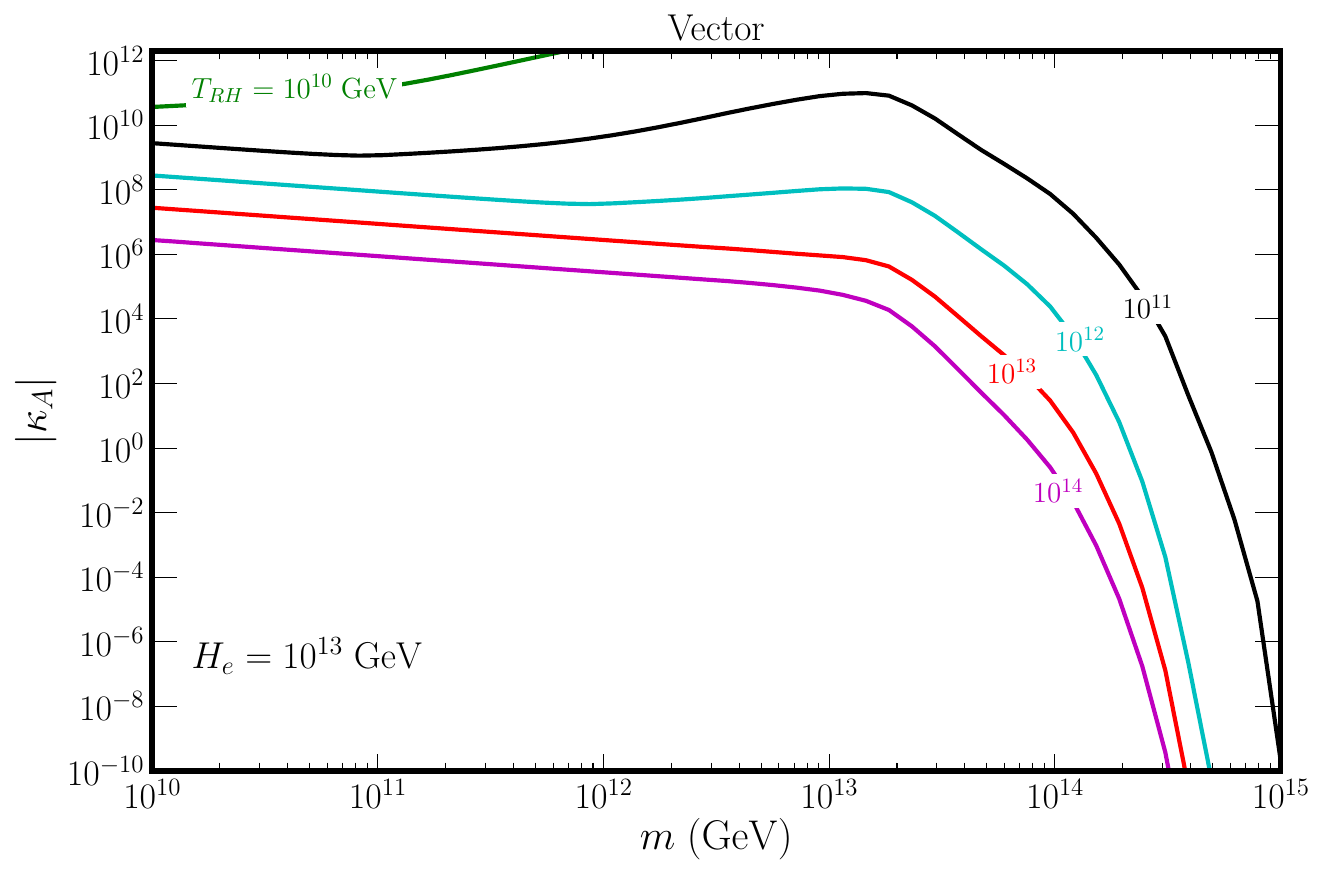}
\caption{\label{fig:vector_density} Same as \fref{fig:fermion_density}, but for the vector WIMPzilla model.}
\end{center}
\end{figure}

If the WIMPzilla is a spin-1 vector $A$, then the coupling of $A$ to the Higgs field, given in \eref{eq:interactions}, is specified by the ratio $\kappa_A m^2/\Mpl^2$.  We evaluate the thermally averaged WIMPzilla annihilation cross section in \aref{app:cross_section} finding 
\begin{align}\label{eq:vector_sigma}
	\langle \sigma v \rangle & = 
	\frac{|\kappa_A|^2}{2592 \pi} \frac{T^2}{\Mpl^4} \left[
	6 \frac{m^2}{T^2} \, \frac{K_1^2(m/T)}{K_2^2(m/T)} 
	+ \frac{4 \sqrt{\pi}}{K_2^2(m/T)}G^{30}_{13}\left( \frac{m^2}{T^2} \left| \begin{array}{l} -1/2 \\ -2, 1, 2 \end{array}\right. \right) \right. \nn & \hspace{36pt}
	\left. - \frac{4 \sqrt{\pi}}{K_2^2(m/T)} G^{30}_{13}\left( \frac{m^2}{T^2} \left| \begin{array}{l} 1/2 \\ -1, 1, 2 \end{array} \right. \right) 	\right] 
	\per
\end{align}
The source term is calculated using \erefs{eq:neq_def}{eq:source_def} with $g =3$, and we obtain 
\begin{align}\label{eq:vector_source}
	\Scal 
	& = 
	\frac{|\kappa_A|^2}{256 \pi^5}\frac{m^4}{\Mpl^4}  T^4 \left[
	6 \frac{m^2}{T^2} \, K_1^2(m/T)
	+ 4 \sqrt{\pi} \, G^{30}_{13}\left( \frac{m^2}{T^2} \left| \begin{array}{l} -1/2 \\ -2, 1, 2 \end{array} \right. \right) \right. \nn
	& \hspace{36pt} \left. - 4 \sqrt{\pi} \, G^{30}_{13}\left( \frac{m^2}{T^2} \left| \begin{array}{l} 1/2 \\ -1, 1, 2 \end{array} \right. \right) 
	\right] 
	\com
\end{align}
We evaluate the integral in \eref{eq:formal_soln_a} to obtain the comoving number density of thermally produced WIMPzilla particles, and the result is shown as the blue curves in \eref{fig:vector_density}.  In the limits of large and small WIMPzilla mass, the density can be approximated as 
\begin{align}\label{eq:vector_nX}
	\frac{a^3 n}{a_e^3H_e^3} \approx 
	\begin{cases}
	\dfrac{33885 |\kappa_A|^2}{8192 \pi^4} \dfrac{T_{\rm max}^{12}}{H_e^4 \Mpl^4 m^4} e^{-2m/T_{\rm max}} \, f_1(m/T_{\rm max}) & \textrm{for} \ T_{\RH} \ll m \\ 
	\dfrac{3|\kappa_A|^2}{8\pi^5} \dfrac{T_{\rm max}^{12}}{H_e^4 \Mpl^4 T_\RH^4} & \textrm{for} \ m \ll T_{\RH}
	\end{cases}
\end{align}
where $f_1(x) \equiv 1 + 2x + 2 x^2 + 4x^3/3 + 2x^4/3 + 4x^5/15 + 4x^6/45 + 32x^7/1255  + 128x^8/33885$.  

%=============

In the left panel of \fref{fig:vector_density} we plot the thermally produced comoving density.  In the right panel we compare the gravitationally produced WIMPzilla abundance to the thermally produced WIMPzilla abundance, and obtain a value of $|\kappa_A|$ where the two sources of WIMPzillas will result in equal abundances.   For $\left|\kappa_A\right| > \left|\kappa_{A*}\right|$, thermal production dominates, while for $\left|\kappa_A\right|< \left|\kappa_{A*}\right|$, gravitational production dominates.   Large values of $|\kappa_A|$ correspond to lowering the cutoff to $\Mpl / \sqrt{|\kappa_A|}$.  As we discussed before \eref{eq:kappamax}, the validity of the EFT requires $\kappa_A \ll 10^{12}$, which is satisfied across \fref{fig:vector_density}.

%==================================
% Thermal WIMPzillas as DM
\section{Dark Matter Produced Through the Higgs Portal}\label{sec:ThermalDM}
%==================================

If the WIMPzilla is stable, as we have assumed, then a relic abundance of WIMPzilla particles will persist in the universe today.  In this section, we assess the region of parameter space in which the WIMPzilla saturates the present dark-matter density.  We also show the regions of parameter space where models are \textit{disallowed} because of overproduction of dark matter.

The WIMPzilla relic abundance today (time $t=t_0$) is given by $\Omega = m n(t_0) / 3 \Mpl^2 H_0^2$ where $H_0 = 100 h \, {\rm km}\ {\rm Mpc}^{-1} {\rm sec}^{-1}$ is the Hubble constant.  Using the $a^{-3}$ scaling behavior for the number density of WIMPzillas, it is straightforward to show that 
\begin{align}\label{eq:Omegah2}
	\Omega h^2 = \bigl( 0.12 \times 10^7 \bigr) \left( \frac{H_e}{10^{13} \GeV} \right)^2 \left( \frac{T_{\RH}}{10^9 \GeV} \right) \left( \frac{m}{H_e} \right) \left( \frac{a^3 n}{a_e^3 H_e^3} \right) \per
\end{align}
The last factor is simply the comoving WIMPzilla number density, which we have calculated in the previous sections.  The dark matter relic abundance is measured to be $\Omega_\DM h^2 \simeq 0.12$.  

\begin{figure}[t]
\begin{center}
\includegraphics[width=0.49\textwidth]{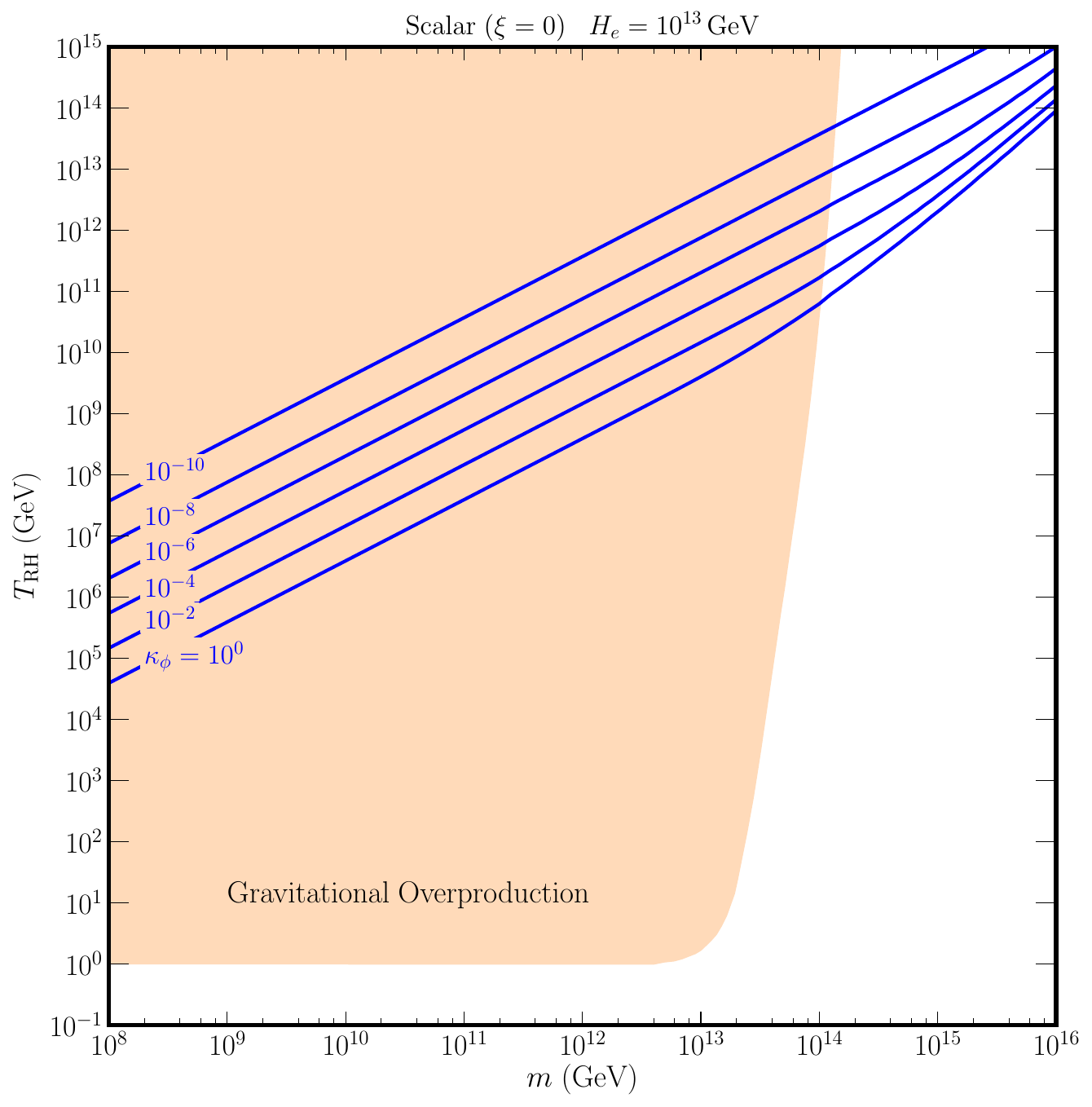} \hfill 
\includegraphics[width=0.49\textwidth]{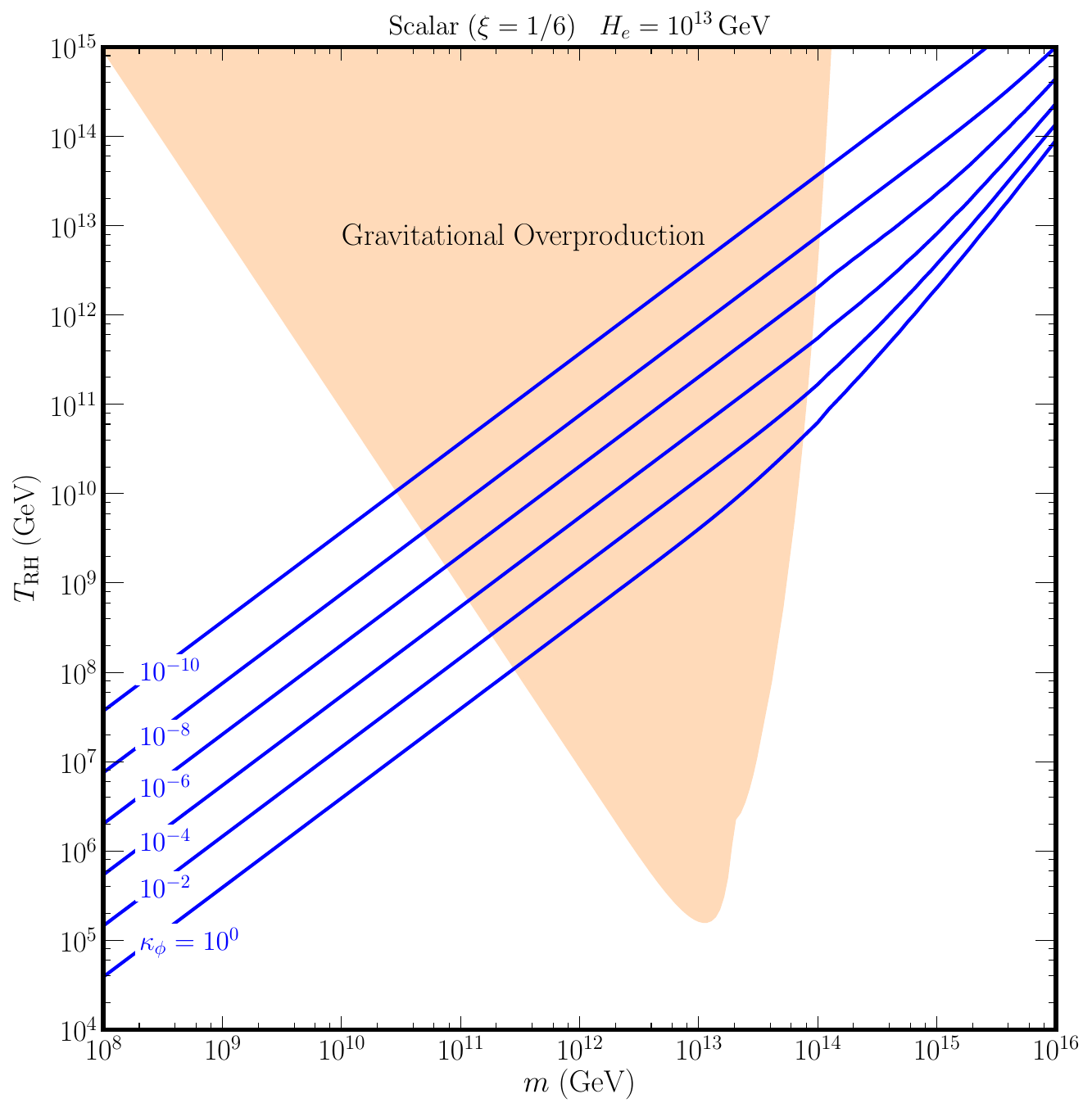} \hfill
\caption{\label{fig:Oh2s}
The region of parameter space where the predicted scalar WIMPzilla abundance (minimal coupling in left panel and conformal coupling in right panel) matches the measured dark matter abundance. See text for explanation.  Constraints on isocurvature perturbations exclude $m/H_e \lesssim 6$ \cite{Chung:2004nh}.}
\end{center}
\end{figure}

\begin{figure}[t]
\begin{center}
\includegraphics[width=0.49\textwidth]{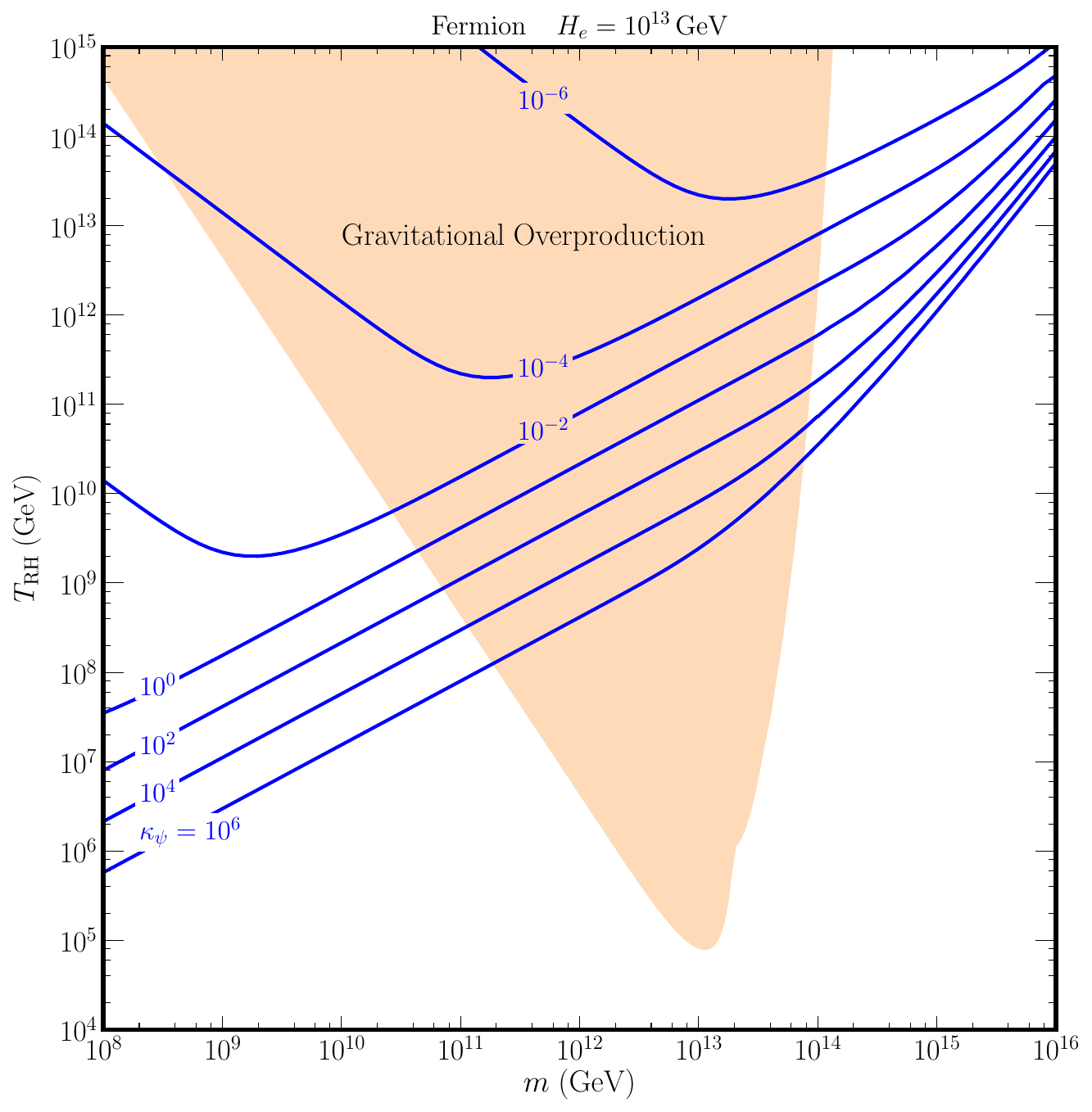} \hfill 
\includegraphics[width=0.49\textwidth]{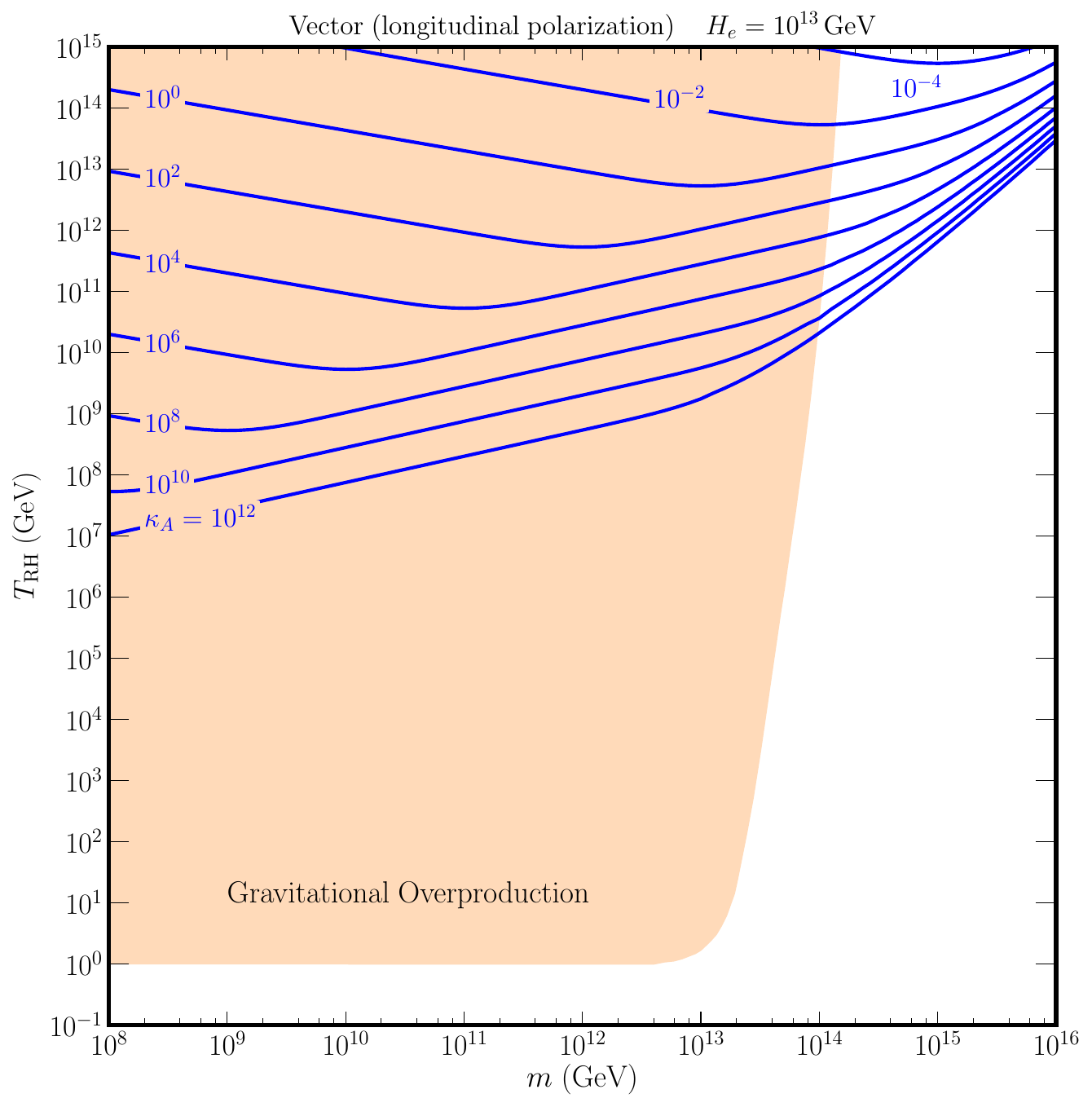}
\caption{\label{fig:Oh2fv}
The region of parameter space where the predicted fermion (left panel) and vector (right panel) WIMPzilla abundance matches the measured dark matter abundance.  See text for explanation.}
\end{center}
\end{figure}

For each of the three models we calculate $\Omega h^2$ using \eref{eq:Omegah2}.  For a given value of the coupling $\kappa$, we determine the values of $m$ and $T_\RH$ that are required to reproduce the present dark matter relic abundance, $\Omega = \Omega_\DM$.  We present our results in Figs.\ \ref{fig:Oh2s} and \ref{fig:Oh2fv}.  The shaded areas in the figures represent regions of parameter space that are \textit{disallowed} because of gravitational overproduction.  (Of course, gravitational production does not depend on $\kappa$.)  Along the edge of the shaded area we obtain $\Omega = \Omega_\DM$ from gravitational particle production alone.  The (blue) curves labeled with values of $|\kappa|$ are the values of $T_\RH$ and $m$ for which thermal production populates WIMPzillas through the Higgs portal in the correct abundance for $\Omega = \Omega_\DM$.  Values of $T_\RH$ \textit{above} the blue curves will result in overproduction of dark matter through thermal processes.  Therefore, the allowed regions of parameter space are outside the shaded area, and below the curves labeled by values of $|\kappa|$.

Values of model parameters $m$, $T_\RH$, and $|\kappa|$ that result in $\Omega = \Omega_\DM$ may be found along the $|\kappa|$ curves that are outside the shaded area, or on the shaded perimeter below the curve corresponding to a given value of $|\kappa|$.

%=============
%=============

%=============
%=============

For the minimally coupled scalar ($\xi=0$) and vector WIMPzilla models, gravitational production is very efficient for $m < H_e$.  The region of parameter space where thermal production can account for all of the dark matter (blue lines) is already excluded by gravitational production (in the shaded area). Conversely, gravitational production is very inefficient for $m > H_e$, and thermal production is the dominant source of WIMPzilla dark matter for $m \gtrsim 10^{14} \GeV$, provided that the reheat temperature is sufficiently large.  For the fermion model and the conformally coupled scalar ($\xi=1/6$) model, gravitational production becomes inefficient for $m < H_e$.  For example, for $|\kappa_\psi| = 1$ and $m \lesssim 10^{11} \GeV$ the dark matter abundance arises primarily from thermal-production.  

We only show values of the couplings $|\kappa|$ that are consistent with the theoretical self-consistency arguments in \eref{eq:kappamax}.  Recall that we have normalized the higher-mass-dimension operators by the Planck mass, and thus $|\kappa| > 1$ implies a lower cutoff $\Lambda \sim \Mpl / |\kappa|$ for the fermion model and $\Lambda \sim \Mpl / \sqrt{\kappa}$ for the vector model.  In the scalar model, the thermally produced relic abundance becomes insensitive to the reheat temperature $T_\RH$ at high values of $T_\RH$.  This is because $a^3 n \sim T_\RH^{-1}$ in the limit of high reheat temperature, as we can see from \fref{fig:scalar_density}, and then $\Omega \sim T_\RH^0$ from \eref{eq:Omegah2}.  

In the upper-right corner of Figs.\ \ref{fig:Oh2s} and \ref{fig:Oh2fv}, the spacing between the blue thermal-production curves begins to shrink.  This is because the thermal abundance becomes exponentially suppressed if the WIMPzilla mass is too large; see Eqs.\ (\ref{eq:scalar_nX}), (\ref{eq:fermion_nX}), and (\ref{eq:vector_nX}).  The Boltzmann suppression sets in where $m \gtrsim \, 10 T_{\rm max}$ with $T_{\rm max}$ given by \eref{eq:Tmax}.  To compensate the exponential suppression, the coupling $|\kappa|$ must be made exponentially large in order for $\Omega h^2$ to match the correct relic abundance.  Hence, $|\kappa|$ changes very rapidly in this regime, and the spacing between the blue curves becomes small.  

Although we have not discussed the power spectrum of dark matter density perturbations here, it is well known that the spectrum is nearly scale invariant for gravitationally produced minimally coupled scalar-WIMPzilla dark matter in the regime $m/H_e \lesssim 1$ \cite{Kuzmin:1998kk}.  Since the dark matter is produced nonthermally, the dark matter density fluctuations are not correlated with the photon density fluctuations, which corresponds to a large dark matter-photon isocurvature.   In fact, minimally coupled scalar-WIMPzilla dark matter is ruled out for $m/H_e \lesssim 6$ from the cosmic microwave background limits on isocurvature \cite{Chung:2004nh}.  The isocurvature constraint does not apply to the models with conformally coupled scalar, fermionic, or vector WIMPzilla dark matter, which have blue power spectra.  

In the regions of parameter space shown in Figs.\ \ref{fig:Oh2s} and \ref{fig:Oh2fv} we have verified that the WIMPzilla does not reach thermal equilibrium by comparing the density $n$ against the would-be equilibrium density $\bar{n}$ at the time when $a^3 n$ becomes constant and freeze-in is completed.  If we were to increase the Higgs-WIMPzilla coupling $|\kappa|$ sufficiently, then the WIMPzilla would thermalize, and its relic abundance would be determined instead by {\it thermal freeze-out}.  In this regime, it can also be possible to achieve the correct relic abundance \cite{Giudice:2000ex}, but the predicted abundance depends also on additional dynamics in the dark sector, such as self-interactions \cite{Heikinheimo:2017ofk}, that are not described by the Higgs portal operators in \eref{eq:interactions}.  Thus we do not consider this scenario here.  

Throughout these calculations we have fixed the Hubble parameter at the end of inflation to be $H_e =10^{13} \GeV$.  Consequently, energy conservation puts an upper limit on the reheat temperature, which is $3 \Mpl^2 H_e^2 > (\pi^2/30) g_{\ast} T_\RH^4$ or equivalently $T_\RH \lesssim 3 \times 10^{15} \GeV$ for $g_\ast = 106.75$.  

%==================================
% CONCLUSION
%==================================
\section{Conclusion}\label{sec:Conclusion}

In this article, we have studied superheavy (WIMPzilla) dark matter in the context of effective field theory.  In previous studies of WIMPzilla production it was customary to assume that the WIMPzilla is noninteracting apart from its coupling with gravity.  However, when the WIMPzilla is viewed from the perspective of effective field theory, one expects additional interactions to arise.  In particular, there is no symmetry to forbid a direct interaction between the WIMPzilla and the Standard Model Higgs field since this interaction is just a product of the WIMPzilla and Higgs mass terms.  The strength of this interaction may be Planck-suppressed (or smaller), but in general one expects it to be present.  In this work, we have studied the role of the direct Higgs-WIMPzilla coupling in the thermal production (freeze-in) of WIMPzilla dark matter.  

The primary new calculation in this work is the derivation of the comoving number density of WIMPzilla dark matter produced from the annihilation of Higgs-boson pairs in the plasma via the interactions in \eref{eq:interactions}.  The numerical results appear in Figs.~\ref{fig:scalar_density},~\ref{fig:fermion_density},~and~\ref{fig:vector_density} for the scalar, fermion, and vector WIMPzilla models respectively, and the corresponding analytic approximations can be found in Eqs.~(\ref{eq:scalar_nX}),~(\ref{eq:fermion_nX}),~and~(\ref{eq:vector_nX}).  By comparing with the abundance of gravitationally produced WIMPzilla dark matter, which was calculated in other works and summarized in \fref{fig:n_grav}, we determined the strength of the Higgs-WIMPzilla coupling at which the thermally produced abundance becomes dominant.  These results appear in Figs.~\ref{fig:kappa*s},~\ref{fig:fermion_density},~and~\ref{fig:vector_density}.  For instance, in the fermion WIMPzilla model we found that even a Planck-suppressed Higgs-WIMPzilla interaction (corresponding to $|\kappa_\psi| = 1$), can be sufficient for the thermal abundance to dominate over the gravitationally produced population if $m \sim T_\RH < H_e$.  

This study leaves open various directions for future work.  For instance, we have focused on a chaotic model of inflation (quadratic inflaton potential) followed by a period of perturbative reheating with an effective equation of state $w_\RH = 0$.  These assumptions could be generalized to consider different models of inflation and reheating.  However, we do not expect that these modifications would change our general conclusions.  It would also be interesting to explore more carefully the gravitational particle production for higher spin fields, such as spin-1 and spin-3/2.

%----------------------------------------------------------------
% Acknowledgements
%----------------------------------------------------------------
\quad \\
\noindent
{\bf Acknowledgments:} 
We are grateful to Yann Gouttenoire for pointing out a couple typographical errors that appeared in Sec.~4 of an earlier version of this article.  The work of E.W.K.\ is supported by the U.S. Dept. of Energy contract DE-FG02-13ER41958.  A.J.L. is supported at the University of Chicago by the Kavli Institute for Cosmological Physics through grant NSF PHY-1125897 and an endowment from the Kavli Foundation and its founder Fred Kavli.

%----------------------------------------------------------------
%----------------------------------------------------------------
%----------------------------------------------------------------
\appendix

%==================================
% APPENDIX
%==================================
\section{Cross Section Calculation}\label{app:cross_section}

Here we evaluate the thermally averaged WIMPzilla annihilation cross section for each of the three models.  Let us first introduce some model-independent definitions.  (See \rref{Gondolo:1990dk} for additional details.)  

Let $\Mcal_{XX \to \Phi \bar{\Phi}}({\bm p}_1, s_1; {\bm p}_2, s_2; {\bm p}_{\Phi}; {\bm p}_{\bar{\Phi}})$ denote the scattering amplitude for the annihilation of a WIMPzilla of momentum ${\bm p}_1$ and spin $s_1$ and a second WIMPzilla with momentum ${\bm p}_2$ and spin $s_2$ into a Higgs boson with momentum ${\bm p}_{\Phi}$ and an anti-Higgs with momentum ${\bm p}_{\bar{\Phi}}$.  There are two annihilation channels [see \eref{eq:two_channels}], and $\Mcal_{XX \to \Phi \bar{\Phi}}$ denotes the matrix element for either one or the other.  Due to the isospin symmetry, these two matrix elements are equivalent, and the final thermally averaged cross section is doubled.  We include this factor of $2$ at the end of the calculation.  

The thermally averaged annihilation cross section is defined by 
\begin{align}\label{eq:sv_def}
	\langle \sigma v \rangle_{XX \to \Phi \bar{\Phi}} 
	& \equiv \frac{1}{\bar{n} \bar{n}} \int \! \frac{\ud^3 {\bm p}_{\Phi}}{(2\pi)^3} \frac{1}{2E_{\Phi}} \int \! \frac{\ud^3 {\bm p}_{\bar{\Phi}}}{(2\pi)^3} \frac{1}{2E_{\bar{\Phi}}} \int \! \frac{\ud^3 {\bm p}_1}{(2\pi)^3} \frac{1}{2E_1} \int \! \frac{\ud^3 {\bm p}_2}{(2\pi)^3} \frac{1}{2E_2} 
	\nn & \qquad 
	\times (2\pi)^4 \delta^4(p_{\Phi}+p_{\bar{\Phi}}-p_1-p_2)
	\sum_{s_1, s_2} \left| \Mcal_{XX \to \Phi \bar{\Phi}} \right|^2 \ {\rm exp}\left[-(E_1 + E_2)/T \right] \com
\end{align}
where $E = \sqrt{|{\bm p}|^2 + m^2}$, and the physical number density $\bar{n}$ is defined in \eref{eq:neq_def}.  Although we could evaluate this integral directly, it is convenient first to  express the integrand in different terms.  We use the definition 
\begin{align}\label{eq:sigma_def}
\sigma_{XX \to \Phi \bar{\Phi}} 
	& = \frac{1}{4F(p_1, p_2)} \left[ \int \! \frac{\ud^3 {\bm p}_{\Phi}}{(2\pi)^3} \frac{1}{2E_{\Phi}} \right] \left[ \int \frac{\ud^3 {\bm p}_{\bar{\Phi}}}{(2\pi)^3} \frac{1}{2E_{\bar{\Phi}}} \right] \nn & \qquad \times (2\pi)^4 \delta^4(p_{\Phi}+p_{\bar{\Phi}}-p_1-p_2) \left| \Mcal_{XX \to \Phi \bar{\Phi}} \right|^2 \com
\end{align}
where the Lorentz scalar function $F(p_1,p_2)$ is 
\begin{align}\label{eq:F_def}
	 \equiv \sqrt{ (p_1 \cdot p_2)^2 - p_1^2 p_2^2}  \per
\end{align}
Note that $F(p_1, p_2) = (1/2) \sqrt{s} \sqrt{s - 4 m^2}$ where $s = (p_1 + p_2)^2$ is the Mandelstam variable.  

Upon averaging over the $g = (2S + 1)$ possible spin projections of the WIMPzilla particles in the initial state, we obtain the spin-averaged annihilation cross section
\begin{align}\label{eq:sbar_def}
	\overline{\sigma}_{XX \to \Phi \bar{\Phi}} 
	& = \frac{1}{g^2} \sum_{s_1} \ \sum_{s_2} \sigma_{XX \to \Phi \bar{\Phi}} \per
\end{align}
With this notation, the thermally averaged annihilation cross section \pref{eq:sv_def} can be written as 
\begin{align}\label{eq:sigv_2}
	\langle \sigma v \rangle_{XX \to \Phi \bar{\Phi}} 
	& = \frac{g^2}{\bar{n} \bar{n}} \int \! \frac{\ud^3 {\bm p}_1}{(2\pi)^3} \frac{\ud^3 {\bm p}_2}{(2\pi)^3} \, \overline{\sigma}_{XX \to \Phi \bar{\Phi}} \, v_{\Mol}(p_1,p_2) \ {\rm exp}\left[ -(E_1 + E_2)/T \right] \com
\end{align}
where we have defined the M{\o}ller velocity
\begin{align}
	v_{\Mol}(p_1,p_2) \equiv \frac{F(p_{1}, p_{2})}{E_{1} E_{2}} = \sqrt{ \left| {\bm v}_1 - {\bm v}_2 \right|^2 - \left| {\bm v}_1 \times {\bm v}_2 \right|^2 }
\end{align}
with ${\bm v} = {\bm p} / E$.  
As long as $\overline{\sigma}_{XX \to \Phi \bar{\Phi}}$ only depends on $s$, which is the case for the models of interest, the other momentum integrals can be evaluated exactly, leaving \cite{Gondolo:1990dk} %Eq.~(3.6)
\begin{align}\label{eq:sigv_final}
	\langle \sigma v \rangle_{XX \to \Phi \bar{\Phi}} & = 
	\frac{g^2}{\bar{n} \bar{n}} \frac{T}{32 \pi^4} \int_{4 m^2}^{\infty} \! \ud s \, (s-4m^2) \, \sqrt{s} \, K_1(\sqrt{s}/T) \ \overline{\sigma}_{XX \to \Phi \bar{\Phi}}(s)
\end{align}
where $K_n(x)$ is the modified Bessel function of the second kind of order $n$. All that remains is to evaluate the cross section $\overline{\sigma}_{XX \to \Phi \bar{\Phi}}(s)$ for each of the models, and perform the final integral in \eref{eq:sigv_final}.

%========
\subsection{Scalar WIMPzilla}\label{app:scalar_WIMPzilla} 
%========

For the scalar case discussed in \sref{sub:scalar_WIMPzilla}, the matrix element is simply $\Mcal_{XX \to \Phi \bar{\Phi}} = \kappa_\phi$, and using \eref{eq:sigma_def} we write the WIMPzilla annihilation cross section as 
\begin{align}
	\sigma_{XX \to \Phi \bar{\Phi}} 
	& = \frac{|\kappa_\phi|^2}{4F(p_{\X_1}, p_{\X_2})} \frac{(2\pi)^4}{(2\pi)^6} \left[ \int \! \frac{\ud^3 {\bm p}_{\Phi}}{2 E_{\Phi}} \frac{\ud^3 {\bm p}_{\bar{\Phi}}}{2 E_{\bar{\Phi}}} \, \delta(p_{\Phi}+p_{\bar{\Phi}}-p_{\X_1}-p_{\X_2}) \right] 
	\per
\end{align}
The integral was evaluated in \rref{Pietschmann:1974ap}, which gives $(\pi/2) (s - 4 m_{\Phi}^2)^{1/2} / s^{1/2}$.  The spin averaging is trivial since all the particles are scalars.  
Combining the various factors, the annihilation cross section is 
\begin{align}
	\bar{\sigma}_{XX \to \Phi \bar{\Phi}} 
	& = \frac{|\kappa_\phi|^2}{16 \pi} \frac{1}{s} \frac{\sqrt{ s - 4 m_{\Phi}^2}}{\sqrt{s - 4 m^2}}  	\per 
\end{align}
Using \eref{eq:sigv_final} with $g=1$ for a scalar WIMPzilla, we write the thermally averaged annihilation cross section as 
\begin{align}
	\langle \sigma v \rangle_{XX \to \Phi \bar{\Phi}} & = 
	\frac{1}{\bar{n} \bar{n}} \frac{|\kappa_\phi|^2}{16\pi} \frac{T}{32 \pi^4} \int_{4 m^2}^{\infty} \! \ud s \, \sqrt{ s-4m^2 } \, K_1(\sqrt{s}/T) \com
\end{align}
where we have also set $m_{\Phi} = 0$.  The equilibrium number densities are given by \eref{eq:neq_def} with $g = 1$.  Evaluating the integral gives the thermally averaged annihilation cross section  
\begin{align}
	\langle \sigma v \rangle_{XX \to \Phi \bar{\Phi}} & = 
	\frac{1}{m^2} \frac{|\kappa_\phi|^2}{32 \pi} \, \frac{K_1^2(m/T)}{K_2^2(m/T)} \per
\end{align}
To account for the two annihilation channels we multiply by a factor of $2$, which yields the expression in \eref{eq:scalar_sigma}.

%========
\subsection{ Fermion WIMPzilla}\label{app:fermion_WIMPzilla}
%========

For the fermion case discussed in \sref{sub:fermion_WIMPzilla}, the spin-summed, squared matrix element is 
\begin{align}
	\sum_{s_1 , \, s_2} \left| \Mcal_{XX \to \Phi \bar{\Phi}} \right|^2 = \frac{2|\kappa_\psi|^2}{\Mpl^2} \left( s - 4 m^2 \right) 	\per
\end{align}
Combining \erefs{eq:sigma_def}{eq:sbar_def} we write the spin-averaged annihilation cross section as 
\begin{align}
	\overline{\sigma}_{XX \to \Phi \bar{\Phi}} 
	& = \frac{1}{4} \frac{1}{4F(p_1, p_2)} \frac{2 | \kappa_\psi |^2}{\Mpl^2} \left( s - 4 m^2 \right)  \frac{(2\pi)^4}{(2\pi)^6} \, \left[ \int \! \frac{\ud^3 {\bm p}_{\Phi}}{2E_{\Phi}} \frac{\ud^3 {\bm p}_{\bar{\Phi}}}{2 E_{\bar{\Phi}}} \, \delta(p_{\Phi}+p_{\bar{\Phi}}-p_{\X_1}-p_{\X_2}) \right] \com
\end{align}
where we have used $g = 2$ for a Majorana fermion WIMPzilla.  As in the scalar calculation, we use \rref{Pietschmann:1974ap} to evaluate the integral, which gives 
\begin{align}
	\overline{\sigma}_{XX \to \Phi \bar{\Phi}} 
	& = \frac{1}{32\pi} \frac{|\kappa_\psi|^2}{\Mpl^2} \  \frac{1}{s} \ \sqrt{ s - 4 m^2 } \sqrt{s - 4 m_{\Phi}^2}  	\per
\end{align}
Putting this into \eref{eq:sigv_final} lets us write the thermally averaged annihilation cross section as 
\begin{align}
\langle \sigma v \rangle_{XX \to \Phi \bar{\Phi}} & = 
\frac{1}{\bar{n} \bar{n}} 
\frac{1}{256 \pi^5} 
\frac{|\kappa_\psi|^2}{\Mpl^2} 
\, T
\int_{4 m^2}^{\infty} \! \ud s \, \left( s-4m^2 \right)^{3/2} \, K_1(\sqrt{s}/T) \com
\end{align}
where we have neglected the Higgs boson mass.  With a change of variables, we can write 
\begin{align}
	\langle \sigma v \rangle_{XX \to \Phi \bar{\Phi}} & = 
	\frac{1}{\bar{n} \bar{n}} 
	\frac{1}{8 \pi^5} 
	\frac{|\kappa_\psi|^2}{\Mpl^2} 
	\, T^6
	\left[ \frac{1}{16} \int_{2 m / T}^{\infty} \! y \, \ud y \left( y^2 - 4m^2 / T^2 \right)^{3/2} \, K_1(y) \right]   \per
\end{align}
The integral can be evaluated in terms of the Meijer G-function, and the quantity in square brackets equals 
\begin{align}\label{eq:Meijer_G_approx}
	\left[ \cdots \right] 
	= \frac{3 \sqrt{\pi}}{8} G^{30}_{13}\left( \frac{m^2}{T^2} \left| \begin{array}{l} 5/2 \\ 0, 2, 3 \end{array} \right. \right)
	\approx \begin{cases}
	\dfrac{3\pi}{8} \dfrac{m^2}{T^2} \left( 1 + \dfrac{7}{4} \dfrac{T}{m} \cdots \right) e^{-2 m/T} &  \mathrm{for} \ T < m \\[6pt]
	1 - \dfrac{3}{4} \dfrac{m^2}{T^2} + \cdots & \ \mathrm{for} \ m < T \per
	\end{cases}
\end{align}
The double exponential suppression arises because collisions producing a pair of $X$ particles (energy $E = 2 m$) can only occur for $\Phi$ particles deep in the high-energy Boltzmann tail of the phase space distribution function.  Now using the expression for $\bar{n}$ from \eref{eq:neq_def} we have 
\begin{align}
	\langle \sigma v \rangle_{XX \to \Phi \bar{\Phi}} & = 
	\frac{1}{8 \pi}
	\frac{|\kappa_\psi|^2}{\Mpl^2} 
	\frac{T^4}{m^4 K_2^2(m/T)} 
	\left[ \frac{3 \sqrt{\pi}}{8} G^{30}_{13}\left( \frac{m^2}{T^2} \left|\begin{array}{l} 5/2 \\ 0, 2, 3 \end{array} \right. \right) \right] 
	\com
\end{align}
which yields \eref{eq:fermion_sigma} after multiplying by a factor of $2$ to account for the two annihilation channels.

%========
\subsection{Vector WIMPzilla}\label{app:vector_WIMPzilla}
%========

For the vector case discussed in \sref{sub:vector_WIMPzilla}, the spin-summed, squared matrix element is 
\begin{align}
	\sum_{s_1 , \, s_2} \left| \Mcal_{XX \to \Phi \bar{\Phi}} \right|^2 = \frac{|\kappa_A|^2 m^4}{\Mpl^4} \left[ 2 + \frac{\left( s - 2 m^2 \right)^2}{4 m^4} \right] 
	\per
\end{align}
The first term in square brackets corresponds to the two transverse polarization states, and the other term corresponds to the longitudinal polarization.  Note that the matrix element diverges in the limit $s/m^2 \to \infty$, which signals a loss of perturbative unitarity.  
As with longitudinal $W$-boson scattering in the SM, perturbative unitarity is regained if the theory is Higgsed in the UV.  Since we will be considering energies as high as $s \sim T_{\rm max}^2$, the validity of our calculation requires the symmetry-breaking scale to be larger than $T_{\rm max} / 4\pi$.  

Combining \erefs{eq:sigma_def}{eq:sbar_def} we write the spin-averaged annihilation cross section as 
\begin{align}
	\overline{\sigma}_{XX \to \Phi \bar{\Phi}} 
	& = \frac{1}{9} \frac{1}{4F(p_1, p_2)} \frac{|\kappa_A|^2 m^4}{\Mpl^4} \left[ 2 + \frac{\left( s - 2 m^2 \right)^2}{4 m^4} \right] \frac{(2\pi)^4}{(2\pi)^6} \nn & \qquad \times \left[ \int \! \frac{\ud^3 {\bm p}_{\Phi}}{2E_{\Phi}} \frac{\ud^3 {\bm p}_{\bar{\Phi}}}{2 E_{\bar{\Phi}}} \, \delta(p_{\Phi}+p_{\bar{\Phi}}-p_1-p_2) \right] \com
\end{align}
where we have used $g = 3$ for a vector WIMPzilla.  This is the same integral that we encountered in the previous subsections, and upon evaluating it we obtain
\begin{align}
	\overline{\sigma}_{XX \to \Phi \bar{\Phi}} 
	& = \frac{1}{9} \frac{|\kappa_A|^2}{16\pi} \frac{m^4}{\Mpl^4} \left[ 2 + \frac{\left( s - 2 m^2 \right)^2}{4 m^4} \right]  \, \frac{1}{s} \frac{\sqrt{s - 4 m_{\Phi}^2}}{\sqrt{s - 4 m^2}} 	\per
\end{align}
Putting this into \eref{eq:sigv_final} lets us write the thermally averaged annihilation cross section as 
\begin{align}
	\langle \sigma v \rangle_{XX \to \Phi \bar{\Phi}} & = 
	\frac{1}{\bar{n} \bar{n}} \frac{m^4}{\Mpl^4} \frac{|\kappa_A|^2}{512 \pi^5} T \, \int_{4 m^2}^{\infty} \! \ud s \, \sqrt{ s-4m^2 } \, K_1(\sqrt{s}/T) \ \left[ 2 + \frac{\left( s - 2 m^2 \right)^2}{4 m^4} \right] \com
\end{align}
where we have neglected the Higgs boson mass ($m_{\Phi} \ll m$).  A change of variables results in 
\begin{align}
	\langle \sigma v \rangle_{XX \to \Phi \bar{\Phi}} & = 
	\frac{1}{\bar{n} \bar{n}} \frac{m^4}{\Mpl^4} \frac{|\kappa_A|^2}{256 \pi^5} T^4 \Biggl\{ \int_{2 m/T}^{\infty} \! \ud y \, y \, \sqrt{ y^2-4m^2/T^2 } \, K_1(y) \nn & \qquad \times \left[ 2 + \frac{\left( y^2 - 2 m^2/T^2 \right)^2}{4 m^4/T^4} \right] \Biggr\} \per
\end{align}
The integral in $\{\cdots\}$ brackets evaluates to 
\begin{align}
	\{\cdots\}
	& = 
	6 \frac{m^2}{T^2} \,  K_1^2(m/T)
	+ 4 \sqrt{\pi} \, G^{30}_{13}\left( \frac{m^2}{T^2} \left| \begin{array}{l} -1/2 \\ -2, 1, 2 \end{array} \right. \right) 
	- 4 \sqrt{\pi} \, G^{30}_{13}\left( \frac{m^2}{T^2} \left| \begin{array}{l} 1/2 \\ -1, 1, 2 \end{array} \right. \right) 
	\nn
	& \approx \begin{cases}
	3 \pi \ \dfrac{m}{T} \left( 1 + \dfrac{11}{4} \dfrac{T}{m} \cdots \right) e^{-2 m/T} &  \ \mathrm{for}\ T < m \\[6pt]
	96 \dfrac{T^4}{m^4} - 24 \dfrac{T^2}{m^2} + 9 + \cdots & \ \mathrm{for}\ m < T \per
	\end{cases}
\end{align}
Now using the expression for $\bar{n}$ from \eref{eq:neq_def} with $g = 3$, we have 
\begin{align}
	\langle \sigma v \rangle_{XX \to \Phi \bar{\Phi}} & = 
	\frac{T^2}{\Mpl^4} \frac{|\kappa_A|^2}{5184 \pi} \left[
	6 \frac{m^2}{T^2} \, \frac{K_1^2(m/T)}{K_2^2(m/T)}
	+ 4 \sqrt{\pi} \, K_2^{-2}(m/T) G^{30}_{13}\left( \frac{m^2}{T^2} \left| \begin{array}{l} -1/2 \\ -2, 1, 2 \end{array} \right. \right) \right. \nn
	& \qquad \left. - 4 \sqrt{\pi} \,  K_2^{-2}(m/T) G^{30}_{13}\left( \frac{m^2}{T^2} \left| \begin{array}{l} 1/2 \\ -1, 1, 2 \end{array} \right. \right)	\right] 
	\com
\end{align}
which yields \eref{eq:vector_sigma} after multiplying by a factor of $2$ to account for the two annihilation channels.  

%----------------------------------------------------------------
% References
%----------------------------------------------------------------
%\bibliographystyle{h-physrev5}
\bibliographystyle{JHEP.bst}
%\billiographystyle{JCAP.bst}
\bibliography{refs--Grav_Part_Prod}

\end{document}